\documentclass[pra,preprint,tightenlines,eqsecnum,nofootinbib]{revtex4}

 \usepackage{amsmath,graphicx}
  
 \newcommand*{\pd}[2]{\frac{\partial #1}{\partial #2}}
 \newcommand*{\od}[2]{\frac{d #1}{d #2}}
 \newcommand*{\x}{\mathbf{x}_\bot}
 \newcommand*{\y}{\mathbf{y}_\bot}
 \newcommand*{\erfc}{\mathop{\textrm{erfc}}\nolimits}
 \newcommand*{\Ei}{\mathop{\textrm{Ei}}\nolimits}
 \newcommand*{\IM}{\mathop{\textrm{Im}}\nolimits}
 \newcommand*{\RE}{\mathop{\textrm{Re}}\nolimits}
 \newcommand*{\length}{[\textrm{length}]}
 \newcommand*{\s}{\scriptstyle }

 \newcommand*{\zb}{\beta}
 \newcommand*{\zg}{\gamma}
 \newcommand*{\zd}{\delta} \newcommand*{\zD}{\Delta}
 \newcommand*{\ze}{\epsilon}
 \newcommand*{\zy}{\theta}
 \newcommand*{\zk}{\kappa}
 \newcommand*{\zp}{\pi}
 \newcommand*{\zr}{\rho}
 \newcommand*{\zt}{\tau}
 \newcommand*{\zf}{\phi}
 
 \newcommand*{\zw}{\omega}
 
 \newcommand*{\zj}{\quad}   \newcommand*{\zJ}{\qquad}  
 \newcommand*{\zH}{\equiv}
 \newcommand*{\zI}{\infty}
 \newcommand*{\zM}{\times}
 \newcommand*{\zV}{\nabla}

 \newtheorem{lemma}{Lemma}
 \newtheorem{theorem}{Theorem}

 \DeclareFixedFont{\fiverm}{OT1}{cmr}{m}{n}{5pt}
 \input{prepictex}
 \input{pictex}
 \input{postpictex}

\begin{document}

 \title{The Dirichlet-to-Robin Transform}

 \author{J. D. Bondurant}
 \altaffiliation{Current affiliation:
  Systat Software, Inc., Richmond, California}
 \author{S. A. Fulling}
 \email{fulling@math.tamu.edu}
 \homepage{http://www.math.tamu.edu/~fulling}
\affiliation{Mathematics and Physics Departments,%
  Texas A\&M University, College Station, TX, 77843-3368 USA}

\date{August 29, 2004}

  \begin{abstract}
    A simple transformation converts a solution of a partial 
    differential equation  with a 
Dirichlet boundary condition to a function satisfying a Robin 
(generalized Neumann) condition.  In the simplest cases this 
observation enables the exact construction of the Green functions 
for the wave, heat, and Schr\"odinger problems with a Robin 
boundary condition.  The resulting physical picture is that the 
field can exchange energy with the boundary, and a delayed 
reflection from the boundary results.  In more general situations
the method allows at least approximate and local construction of 
the appropriate reflected solutions, and hence a ``classical path''
analysis of the Green functions and the associated spectral 
information. 
 By this method we solve the wave equation on an interval with one 
Robin and one Dirichlet endpoint, and thence derive several
 variants of a
 Gutzwiller-type expansion for the density of eigenvalues.
 The variants are consistent except for
 an interesting subtlety of distributional convergence
 that affects only the neighborhood of zero in the frequency 
variable.
 \end{abstract}

 \maketitle

 \section{\label{sec:intro}Introduction}

 Here we develop a technique for constructing a solution to a
differential equation with a Robin (generalized Neumann) boundary 
condition when a solution to the same or a related equation with 
the Dirichlet boundary condition is available.
 The idea is surely not new.
 (In fact, our key formula (\ref{RT}) for the heat equation was 
published in 1891 \cite{Bryan1},
and 
Tikhonov and Samarsky 
\cite{TS} 
 indicate how to solve the wave 
equation on the half-line by a method related to ours.
In both cases the initial heuristic motivation is somewhat 
different from ours, but the resulting calculus is the same.)
 But we believe that  it has not heretofore been systematically 
developed and exploited.

 We are primarily interested in the integral kernels (Green functions) 
 that solve the wave, heat, Schr\"odinger, 
\dots\ equations associated with a given self-adjoint (usually 
positive) differential operator, say~$H$, in the spatial variables.
 These functions are useful not only to solve the partial 
differential equation concerned, but also to obtain information 
about the spectrum and eigenfunctions of~$H$.
 In particular, semiclassical approximation (or classical-quantum 
duality \cite{CPS}) relates the eigenvalues of~$H$ (collectively)
 to the periodic orbits of the classical system whose quantum 
Hamiltonian is $H$ (or of the geometrical optics of the wave 
equation of~$H$).

In the simplest cases 
 (the Laplacian operator with zero-dimensional or at least flat boundaries)
 exact solutions of the Robin problems can be found.
 These kernels manifest a certain nonlocal behavior in the time 
variable, which is reflected in the more familiar eigenfunctions 
and frequency-domain Green functions by a nonpolynomial dependence 
of the reflection coefficients on frequency ($\zw$ in (\ref{eig}), for 
example). This property makes it difficult to solve the 
 time-dependent problems by simple matching at the boundary --- 
 hence the utility of the Dirichlet-to-Robin transform technique.
The phenomenon of time delay at the boundary can be understood 
physically by observing \cite{TS,CZ} that in a wave equation the 
Robin 
condition models an \emph{elastic support} at the boundary;
 in other words, in one dimension a vibrating string is attached at
one endpoint to a discrete, massless spring with Hooke 
constant~$\zk$.
 The string can exchange energy with the spring; this explains the 
surface energy and action associated with Robin boundaries in 
quantum field theory \cite{KCD,RS,systemat,dAC} and also the 
possibility of delayed reflection of a physical impulse. 

 After setting up the general formalism of the method in 
Sec.~\ref{sec:basic},
 in Sec.~\ref{sec:elem} we construct the Green functions for a 
variety of 
 time-dependent problems on the half-line.
 For the heat and Schr\"odinger equations the extension to a 
 higher-dimensional flat boundary is easy, and the analysis leads 
to a (possibly new) determination of the heat-kernel coefficient
(e.g., \cite{BG}) associated with $\zk^n$ for any~$n$.
 Sec.~\ref{sec:adv} discusses the hopes for extending the 
method to less elementary models, where only approximate solutions 
can be expected. 
 Sec.~\ref{sec:interval} treats the wave equation on an interval 
with one Robin endpoint; therefrom, 
 the eigenvalues and the local spectral density are recovered  
 from sums over the periodic and closed orbits of the problem.
 In fact, we stress that there are quite a few different ways of 
arriving at such sums and it is not always obvious that the 
results are the same.
Most notably,
 because the series are not absolutely convergent, their behavior 
is sensitive to reordering of the terms;
we show, however, that this problem is significant in practice only 
at very small frequencies 
 and amounts in principle to a delta-function ambiguity at zero 
frequency (see Appendix~\ref{app:delta}).
 This observation probably has implications for more general 
problems.
 Apart from that phenomenon, we demonstrate agreement among several 
variants of the periodic-orbit expansion.
 A later paper \cite{Fprog} will derive the Casimir energy for a Robin 
 plate  \cite{RS} by the frequency-cutoff method 
 (cf.\ Sec.~\ref{ssec:lap} and \cite{systemat,funorman}).

\goodbreak
\subsubsection*{Notational remarks:}

\begin{enumerate}

 \item For a half-space $\{x>0\}$ we write a Robin boundary 
condition at $x=0$ as 
 \begin{equation}
 \pd ux(0) = +\zk u(0). 
 \label {rob} \end{equation}
Note that $\pd{}x$ is the inward normal derivative.
 Thus $\zk$ equals $-\zg$ in the notation of \cite{systemat}, 
 $-\zb^{-1}$ in the notation of \cite{RS}, 
  $ -S$ in the notation of \cite{BG},
 and $+\zk$ in the notation of \cite{BB1,lgacee3,SPSUS}.
 With this sign convention, $\zk\ge0$ is the physically more
plausible case.  
 (In a heat conduction problem, for example, $\zk<0$ represents 
heat flow from a colder medium to a hotter one.)
 When $\zk\ge0$ all the operators studied in the present paper have 
nonnegative spectrum.

 \item Like \cite{RS} we use $\x$ to stand for ``irrelevant 
transverse dimensions'' although the notation $\mathbf{x}_\|$ 
 (for ``dimensions parallel to the boundary'') would be equally 
logical.

  \item In this paper we find it convenient to use $G$ as a 
  generic notation for all Green functions, rather than introduce 
  separate letters for heat kernels, wave propagators, etc.
 The meaning of $G$ is stable within each subsection.

 \item When a Dirac delta function appears at an endpoint of an 
interval of integration, its interpretation is ambiguous by a 
factor~$\frac12$.
 We adopt the convention that the density of (nonnegative) 
 eigenvalues, $E_k = \zw_k{}\!^2$, with 
respect to integration over the eigenfrequency $\zw\equiv\sqrt{E}$,
 is
 \equation
  \zr(\zw) = \sum_{k=1}^\zI \zd(\zw-\zw_k)  \label{rhodef}
 \endequation
 \emph{even when $0$  is an eigenvalue}.
 The reader may note (especially in Sec.~\ref{ssec:neum})
  some factors $\frac12$ 
that are not fully explained but are clearly necessary to produce 
the right answer.  A rigorous treatment of such issues may appear 
elsewhere \cite{EFprog}.

 \item All equations are in natural units, where time has units 
of $\length^2$ in heat and quantum problems but of $\length$ in wave 
and vacuum-energy problems.
 The only ``coupling constant'' that appears is~$\zk$, with units
 $\length^{-1}$.
 The only dimensionless small parameters are ratios of the lengths
 $\zk^{-1}$, $t$ (or $\zw^{-1}$), and 
 (in Sec.~\ref{sec:interval})~$L$.
Introduction of  $\hbar$ into this context would merely be 
obfuscatory.

 \item We count dimensions nonrelativistically (e.g., a 
vibrating string obeys the \emph{one}-dimen\-sional wave equation).

 \end{enumerate}

 \section{\label{sec:basic}A Dirichlet-to-Robin mapping}

 \subsection{\label{ssec:RtoD}The Robin-to-Dirichlet map, $T$}

 Consider functions $f(x,\x)$ defined on the half-space 
$\mathbf{R}_+^d \zH \{(x,\x) \in \mathbf{R}^d \colon x>0, \, \x 
\in \mathbf{R}^{d-1}\}$
 and satisfying reasonable technical conditions 
  (made more precise below).
 We define (for a given constant $\kappa \in \mathbf{R}$)
 \begin{equation}
 Tf \zH \pd fx - \kappa f.
 \label{Toper}\end{equation}
 Our first lemma is  a tautology, but will prove to be 
powerful:

 \begin{lemma} \label{lem:taut}
 $f$ satisfies the Robin boundary condition (\ref{rob}) at $x=0$
 if and only if $Tf$
 satisfies the Dirichlet condition, $Tf(0,\x)=0$.
 \end{lemma}
 
 Now allow $f$ to depend on an additional variable, $t$,
 and consider differential equations of the general form
\begin{equation}
 \zV^2 f \zH \pd{^2f}{x^2} + \zV^2_\bot f =\zd f,
 \label{pdeform}\end{equation}
 where the operator $\delta$ does not involve $(x,\mathbf{x}_\bot)$
 or derivatives with respect to them;
 for example, $\zd=\partial/\partial t$ gives the heat equation, and
 $\zd=0$ gives the $d$-dimensional Laplace's equation.
 Let $\mathbf{S}$ stand for the natural domain of each problem:
$\mathbf{R}_+^d$ for Laplace's equation or the eigenvalue problem 
 (Sec.~\ref{ssec:eigen}),
    $\mathbf{R}_+^{d+1}$ for the wave equation, and 
 $\mathbf{R}_{++}^{d+1}$ in problems where $t$ is inherently positive
(Secs.~\ref{ssec:heat} and~\ref{ssec:lap}).

 \begin{lemma}\label{lem:tmap}
 If $f$ solves 
 $\zV^2 f= \zd f$
 in $\mathbf{S}$, then so does $Tf$.
\end{lemma}

  Lemma \ref{lem:tmap} follows immediately from the commutativity 
  of $T$ with 
$\zV^2$ and $\zd$.  
 It therefore hinges on the facts that $\zV^2$ has 
constant coefficients, $\kappa$ is a constant function, and the 
boundary is flat 
 (so that the normal differentiation is $\partial/\partial x$).
 Nevertheless, we expect our ultimate construction to be 
useful in more general problems, as explained in Sec.~\ref{sec:adv}.

 \subsection{\label{ssec:backpos}
 Construction of $T^{-1}$ for $\kappa >0$}

 Given a function $g$, we wish to construct an $f$ such that 
$g=Tf$.
 To make $f$ unique we must impose a supplementary condition.

 The differential equation to be solved 
 (with $\x$ suppressed for notational simplicity)
 is $f'(x) -\kappa f(x) = g(x)$,
 whose general solution is
 \begin{equation}
 f(x) = e^{\kappa x} \int_0^x e^{-\kappa s} g(s)\, ds
 + C e^{\kappa x}. 
 \label{gensol}\end{equation}
 It is natural to choose the solution with minimal growth as
  $x\to \infty$. 
  If $g$ obeys a reasonable growth condition 
 (such as boundedness), 
 then one can set
 \begin{equation}
 C = - \int_0^\zI  e^{-\kappa s} g(s)\, ds
 \label{constC}\end{equation}
to cancel the  exponentials and get a solution of similarly 
reasonable growth:
\begin{eqnarray}
  f(x,\x) &\zH& T^{-1}g =
 - e^{\kappa x} \int_x^\zI e^{-\kappa s} g(s,\x)  \,ds 
 \nonumber\\
 &=& - \int_0^\zI e^{-\kappa \ze} g(\ze+x,\x)  \,d\ze.
 \label{T1}\end{eqnarray}
 That is, if the domain of $T$ is suitably restricted, an inverse 
operator exists and is given by formula (\ref{T1}).
 A short calculation verifies that $T^{-1}$ commutes with $\zV^2$,
 and so we have a converse to Lemma~\ref{lem:tmap}:

 \begin{lemma}\label{lem:tback}
 If $g$ solves      $\zV^2 f= \zd f$
 in $\mathbf{S}$, 
  then so does $T^{-1}g$.
\end{lemma}

  The condition $g(0,\x)=0$ is not used in the proofs
of Lemmas \ref{lem:tmap} and~\ref{lem:tback}.
 But now put them together with Lemma~\ref{lem:taut}:

 \begin{theorem}\label{th:transform}
 If $g$ solves the Dirichlet problem for $\zV^2 f= \zd f$
 in $\mathbf{S}$, then $T^{-1} g$ solves the 
corresponding Robin problem (with the given $\zk$), and vice versa.
 \end{theorem}

 \subsection{\label{ssec:backneg}
 Construction of $T^{-1}$ for $\kappa <0$}

This time it is convenient to treat the case $n=1$ thoroughly 
before introducing the complication of transverse dimensions.
 In that case $y_*(x) \zH e^{\zk x}$ is a normalizable 
eigenfunction of the problem; it satisfies $Ty_*=0$, and therefore
 $T$ can't be invertible.
The most convenient growth condition is to require the functions to 
be square-integrable, so that a generalized inverse can be defined 
in the Hilbert-space orthogonal complement of $y_*\,$.

 Suppose, for example, that $u(t,x)$ is to satisfy the heat 
equation with initial data $f(x)$.
 The part of the solution proportional to $y_*$ can be written down 
immediately as    $e^{\zk^2 t} Pf $ where
 \begin{equation}
 Pf = \frac{\langle y_* |f\rangle}{\|y_*\|^2} \, y_*(x)
\label{Pproj}\end{equation}
is the orthogonal projection onto~$y_*\,$. 
Then the full solution is $u(t,x)= u_\bot(t,x) + e^{\zk^2 t} Pf$,
 where $u_\bot(0,x)= (1-P)f$ and $u_\bot(t,x)$ remains orthogonal 
to $y_*$ at all~$t$.
 Our task is just to construct $u_\bot\,$.

 Accordingly, we now treat $T$ as an operator in 
  the Hilbert space $\,\text{range}(1-P)$
 and demand that the solution (\ref{gensol}) lie in this space
 --- i.e., $\langle y_*|f\rangle=0$.
It is easy to see that then (\ref{constC}) is replaced by
 \begin{equation}
 C = - \int_0^\zI  e^{+\kappa s} g(s)\, ds 
 = - \langle y_*|g \rangle,
 \label{constCneg}\end{equation}
and
 \begin{equation}
f(x) \zH T^{-1}g =
  e^{\kappa x} \left [ \int_0^x e^{-\kappa s} g(s)  \,ds
 - \int_0^\zI  e^{+\kappa s} g(s)\, ds \right ].
 \label{1DT2}\end{equation}

 \emph{Remarks:\/} (1) In both cases one can write
 $C = -\int_0^\zI e^{-|\zk|s} g(s)\,ds$.
 (2) Because of the exponential decay of $y_*\,$, (\ref{1DT2})
 makes sense for many functions that are not square-integrable.
 Furthermore, both (\ref{T1}) and (\ref{1DT2}) can be applied to 
certain distributions by duality, and that will be done without 
comment in later sections.

 When $n>1$ the kernel of $T$, as an operator in $L^2(\mathbf{R}_+^d)$,
 consists of products of $e^{\zk x}$ with 
 square-integrable functions of $\x$.
   It is still true that the appropriate formula is
 \begin{equation}
f(x,\x) \zH T^{-1}g =
  e^{\kappa x} \left [ \int_0^x e^{-\kappa s} g(s,\x)  \,ds
 -\int_0^\zI e^{+\zk s}g(s,\x)\,ds \right ],
 \label{T2}\end{equation}
 and that the domain of  $T$ is characterized by the requirement 
that $\int_0^\zI e^{\zk x} f(x,\x)\, dx = 0 $ (for all $\x$).
 This conclusion can be justified by performing a Fourier 
transformation in $\x$ and applying the reasoning above 
  to each Fourier component separately, or 
merely by verifying that (\ref{T2}) satisfies all the necessary 
conditions. 
 Thus Theorem \ref{th:transform}, suitably interpreted,
  applies to both positive and negative~$\zk$.

 \subsection{\label{ssec:eigen}The eigenfunctions}

As a first ``application'' of Theorem \ref{th:transform}, 
 we check that it yields 
the correct eigenfunctions 
 (solutions of $\zV^2 f= -\zw^2 f$)
 of the Robin problem. 
 Ignoring the inert transverse dimensions, we have the Dirichlet 
eigenfunction $g_\zw(x)=\sin (\zw x)$ for each $\zw>0$.
 Applying (\ref{T1}) or (\ref{T2}), as appropriate, gives 
 \begin{equation}
  f_\zw(x) = \frac{-\zw}{\zw^2 +\zk^2} 
 \left [\cos(\zw x) + \frac{\zk}{\zw}\,\sin(\zw x) \right ],
 \label{eignonorm}\end{equation}
 which agrees up to normalization with the Robin eigenfunction
given in standard references \cite{AG,Stak}.
(Of course, this is the hard way to reach an elementary result.)  
In Sec.~\ref{ssec:wave} we shall obtain the normalization by our 
methods as well.

 \section{\label{sec:elem}Applications to elementary Green functions}

 \subsection{\label{ssec:genties}Generalities}

 Henceforth we  restrict attention to $\zk>0$.

 A Green function (integral kernel) associated with a Dirichlet 
problem in $\mathbf{S}$ typically has the ``image charge'' form
 \begin{equation}
 G_D(x,\x,y,\y) = G(x-y,\x-\y) - G(x+y,\x-\y), 
 \label{D}\end{equation}
 where $G(x-y,\x-\y)$ 
 (which is even under interchange of $x$ and $y$
 as well as translation-invariant) 
 is the corresponding Green function for all of $\mathbf{R}^d$
 (and the time variable, if any, is momentarily suppressed in the 
notation). 

 These Green functions represent operators that are functions of 
$\zD$ and hence commute with $T$. 
 Therefore, in operator notation,
 $               G_R \zH  T^{-1}G_D T$
 should be the corresponding operator for the Robin problem.
 It is understood that the action of a Green function on a function 
is
\[
 Gf(x,\x) = \int_0^\zI dy \int_{\mathbf{R}^{d-1}} d\y \,
 G(x,\x,y,\y) f(y,\y).
 \]
 Therefore, the multiplication by $T$ on the right is represented 
by the transpose (real adjoint) of $T$ acting on the variable $y$:
 \[
 T^\dagger_y = -\,\pd{}y -\zk.
 \]
 The multiplication by $T^{-1}$ on the left is represented by the 
corresponding integral operator applied to the $x$ variable of $G$.

 On a function of the form $G(x-y)$, $T^\dagger_y$ 
is equivalent to $T_x\,$.
 On a function of the form $G(x+y)$, passing from $y$ to $x$ leaves 
$T^\dagger$ as $T^\dagger$, which can also be written as
 \begin{equation}
 T^\dagger = -T -2\zk.
 \label{Tdag}\end{equation}
 Since $T$ commutes with $G$, the effect of the similarity 
transformation is to leave the first (direct) term of $G_D$ 
unchanged, while the second (reflected) term changes in a rather 
simple way, stated in a mixed operator/function notation 
 (with transverse variables suppressed) in the next lemma.

\begin{lemma}\label{lem:Tgreen} 
In the context of (\ref{D}),
 \begin{description}
 \item[\rm(a)] \ $T^{-1}G(x-y)T = G(x-y)$,
 \item[\rm (b)] \ $- T^{-1}G(x+y)T = + G(x+y) + 2\zk T^{-1}_x G(x+y)$.
 \end{description}
 \end{lemma}  

 Putting the two parts of the lemma together, we get our principal 
working equations for the rest of this section, 
 (\ref{GR})--(\ref{DeltaG}).

 \begin{theorem}\label{th:green}
 When a Dirichlet Green function has the form (\ref{D}), the 
Green function for the corresponding Neumann problem is
 \begin{equation}
 G_N(x,\x,y,\y) = G(x-y,\x-\y) + G(x+y,\x-\y), 
 \label{N}\end{equation}
 and the one for the Robin problem is
 \begin{equation}
 G_R(x,\x,y,\y) = G_N+  \zD_\zk G
 \label{GR}\end{equation}
where
\begin{equation}     
\zD_\zk G(x,\x,y,\y) = 2\zk T^{-1}_x G(x+y,\x-\y), 
 \label{DeltaG}\end{equation}
 \end{theorem}

 \subsection{\label{ssec:wave}The wave equation in one space dimension}

 Consider the wave problem
 \begin{equation}
 \pd{^2u}{t^2} = \pd{^2u}{x^2}\,, \zJ 
 u(0,x) = f(x), \zj \pd ut(0,x) = 0.
 \label{wavep}\end{equation}
 The well known d'Alembert solution corresponds to the Green 
function 
 \begin{equation}
 G(t,x-y) =\tfrac12 [ \delta(x-y-t) + \delta(x-y+t)].
 \label{dalem}\end{equation}
 From (\ref{D}),  therefore,
 \begin{equation}
 G_D(t,x-y) = \tfrac12 [ \delta(x-y-t) + \delta(x-y+t)]
 -\tfrac12 [ \delta(x+y-t) + \delta(x+y+t)],
 \label{waveD}\end{equation}
 where the last of the four terms is relevant only for $t<0$ since 
$x$ and $y$ are positive in the physical region.
 Thus, by Lemma~\ref{lem:Tgreen}, we are interested in
\[
  -\,\tfrac12 \, T^{-1}\zd(x+y-t)T 
=   \tfrac12 \zd(x+y-t) +\zk T^{-1}_x\zd(x+y-t),
 \]
 \begin{eqnarray*}
  T^{-1}_x\zd(x+y-t) &=& 
  - e^{\kappa x} \int_x^\zI e^{-\kappa s} \zd(s+y-t)  \,ds \\
&=& - e^{\zk (x+y-t)}\zy(t-y-x), 
 \end{eqnarray*}
 where $\zy$ is the unit step function.
 So, finally, for $t>0$ and $\zk>0$ we have
 \begin{equation}
 G_R(t,x,y) = \tfrac12 [ \delta(x-y-t) + \delta(x-y+t) +\zd(x+y-t)]
    -\zk e^{\zk (x+y-t)}\zy(t-x-y).
 \label{waveR}\end{equation}
  
\begin{figure}
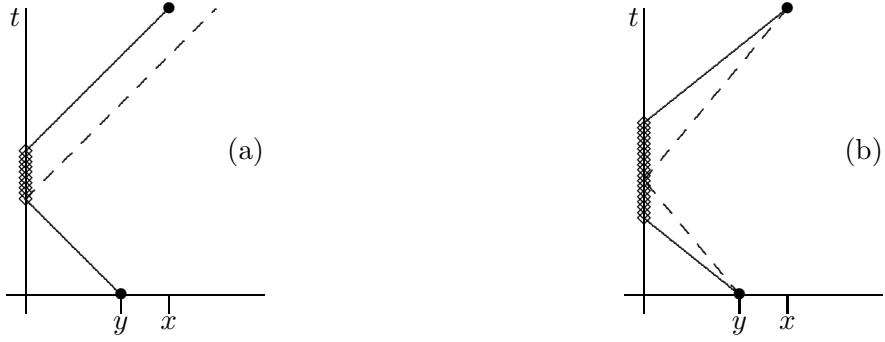

\noindent\null\hfill
 \beginpicture
\setcoordinatesystem units <0.5truein,0.5truein>
  \put{(a)} [r] at 2.5 1.5
 \putrule from 0 -.2 to 0 3
 \putrule from -.2 0 to 2.5 0
 \put{$t$} [rt] <-2pt,0pt> at 0 3
 \put{$\bullet$} at 1 0
 \putrule from 1 0 to 1 -.2
 \put{$y$} [t] <0pt,-1pt> at 1 -.2
 \put{$\bullet$} at 1.5 3
 \putrule from 1.5 0 to 1.5 -.2
 \put{$x$} [t] <0pt,-1pt> at 1.5 -.2
\plot 1 0
      0 1 /
 \plot 0 1.5
       1.5 3 /   
 \setdashes
 \plot 0 1
       2 3 /
\setplotsymbol ({$\diamond$}) \setdots <2pt>
 \plot 0 1
       0 1.5 /
 \endpicture
 \hfill\hfill
 \beginpicture
\setcoordinatesystem units <0.5truein,0.5truein>
 \put{(b)} [r] at 2.5 1.5
 \putrule from 0 -.2 to 0 3
 \putrule from -.2 0 to 2.5 0
 \put{$t$} [rt] <-2pt,0pt> at 0 3
 \put{$\bullet$} at 1 0
 \putrule from 1 0 to 1 -.2
 \put{$y$} [t] <0pt,-1pt> at 1 -.2
 \put{$\bullet$} at 1.5 3
 \putrule from 1.5 0 to 1.5 -.2
 \put{$x$} [t] <0pt,-1pt> at 1.5 -.2
\plot 1 0
      0 .8 /
 \plot 0 1.8
       1.5 3 /   
 \setdashes
\plot 1 0
      0 1.2 /
 \plot 0 1.2
       1.5 3 /
\setplotsymbol ({$\diamond$}) \setdots <2pt>
 \plot 0 .8
       0 1.8 /
 \endpicture\hfill\null
 \caption{(a) For the wave equation with a Robin boundary,
  there is exactly one delayed path (solid curve) 
 from space-time point $(0,y)$
 to a generic point $(t,x)$.
 With a Neumann or Dirichlet boundary, no delay occurs and 
generically the reflected path (dashed curve) misses  $(t,x)$.
(b) For the Schr\"odinger equation, there is exactly one path 
without delay from $(0,y)$ to  $(t,x)$ (dashed curve);
  it has constant speed $v_0\,$. When delayed reflection is 
  possible, there are infinitely many delayed paths (solid curve) 
  with all speeds $v>v_0\,$.}
 \label{fig:paths} \end{figure}

 That is, the solution of the wave equation with Robin boundary 
condition and initial data $u(0,x)=f(x)$, $f(y)$ being interpreted 
as $0$ when $y<0$, is    (for $t>0$)
 \begin{equation}
 u(t,x) = \tfrac12[f(x-t) +f(x+t) + f(t-x)] 
 -\zk \zy(t-x) \int_0^{t-x} e^{-\zk(t-x-y)} f(y)\, dy.
 \label{wavesol}\end{equation}
As usual in one-dimensional wave problems, the solution contains a 
right-moving unreflected pulse, $f(x-t)$, and a left-moving pulse,
 $f(x+t)$, which reflects off the boundary when $t=x$ as a new 
 right-moving pulse, $f(t-x)$.
 In addition, in the Robin problem there is a smeared-out term,
 which is perhaps more illuminatingly rewritten
 \begin{equation}
  -\zk \zy(t-x) \int_0^{t-x}  e^{-\zk \ze} f(t-x-\ze)\, d\ze.
 \label{wavesol2}\end{equation}
Physically, this formula indicates that not all of the pulse is 
reflected immediately --- the wave continues to leak out of the 
boundary with an amplitude that decays exponentially in the 
 time delay $\ze$.
 There is not only an ``echo'' but also ``ringing''\negthinspace.
 Otherwise put:  In the Dirichlet and Neumann problems, the 
signal at $x$ caused by a source at $y$ is a sum over all the 
unit-speed ``classical paths'' from $(0,y)$ to $(t,x)$, including 
possible reflections; there is at most one such path
 (for most points none).
 But in the Robin problem one must also integrate over all the 
paths that travel from $(0,y)$ to the boundary with unit speed,
 stay there for a time $\ze$, and then go with unit speed to $(t,x)$;
 for  $t>x+y$ there is always exactly one such path 
 (Fig.~\ref{fig:paths}(a)).
 Interference between the two terms of (\ref{wavesol}) is 
responsible for conservation of energy.

 Formula (\ref{waveR}) is the central result of this section.
 It can easily be generalized to negative $t$ and also to 
problems with nonzero initial data for $\pd ut$.
 For later reference we note that 
 \begin{equation}
 \int_0^\zI \zD_\zk G(t,x,x)\, dx = -\zk\int_0^\zI e^{\zk(2x-t)} 
\zy(t-2x)\, dx = \tfrac12 (e^{-\zk t}-1). 
 \label{wt}\end{equation}

 Finally, we can recover the eigenfunctions, complete with
 normalization (relative to the Lebesgue measure $d\zw$).
 The generalized eigenfunction expansion
\begin{equation}
  G_R(t,x,y) = \int_0^\zI \cos(\zw t) \psi_\zw(x) \psi_\zw^*(y)
  \, d\zw
 \label{waveeig}\end{equation}
 must hold,
hence
 \begin{equation}
 \psi_\zw(x) \psi_\zw^*(y) = \frac2{\pi} \int_0^\zI G_R(t,x,y)
 \cos (\zw t)\, dt. 
 \label{pk}\end{equation}
 From (\ref{waveR}) one gets for the right-hand side of 
 (\ref{pk})
 \[
 \frac{2/\pi}{\zw^2+\zk^2} [\zw^2 \cos {\zw x} \cos {\zw y}
 +\zk\zw (\sin{\zw x} \cos {\zw y} +\cos {\zw x} \sin {\zw y})
  +\zk^2 \,\sin {\zw x} \sin {\zw y} ],
 \]
 which can be factored as $\psi_\zw(x) \psi_\zw^*(y)$
 with 
 \begin{eqnarray}
  \psi_\zw(x) &=& \sqrt{\frac{2}{\pi}}\, (\zw^2+\zk^2)^{-1/2}
 (\zw \cos {\zw x} + \zk \sin {\zw x}) \label{eig}\\
 &=& \sqrt{\frac2{\pi}} \sin (\zw x +\phi), 
 \nonumber\end{eqnarray}
 where
 \begin{equation}
 \phi \zH \tan^{-1} \frac\zw\zk\,;\zJ
 \sin\zf = \frac\zw{\sqrt{\zw^2+\zk^2}}\,, \zj
 \cos\zf = \frac\zk{\sqrt{\zw^2+\zk^2}}\,. 
 \label{phi}\end{equation}
This normalization agrees with that in the treatises 
\cite{AG,Stak}.

 \subsection{\label{ssec:heat}The heat equation}

 The Green functions for the heat and the Schr\"odinger equation 
are essentially the same algebraically,
  one being an analytic continuation of the other.
 In this subsection we derive the heat kernel by the 
 Dirichlet-to-Robin transformation and compare with the result of 
a direct eigenfunction expansion.
 In the next subsection we recast the Schr\"odinger kernel as a 
``sum over classical paths"\negthinspace.

 The heat kernel for $\mathbf{R}^d$ is
 \begin{equation}
 G(t,x,\x,y,\y) = \frac1{(4\pi t)^{d/2}} 
 \exp \left[-\,\frac{(x-y)^2+(\x-\y)^2}{4t}\right].
 \label{heatker}\end{equation}
The heat kernels for the Dirichlet and Neumann problems
  in the $d$-dimensional 
half-space are then given by (\ref{D}) and (\ref{N}).
 These functions are products of the corresponding one-dimensional 
functions of $x$ and $y$ by the free $(d-1)$-dimensional heat kernel,
 which is unaffected by the operations $T$ and $T^{-1}$
  (for constant~$\zk$). 
 Therefore, 
in what follows we look only at the one-dimensional kernel to 
streamline the notation.

 According to Theorem \ref{th:green},
  the Robin heat kernel equals the Neumann kernel, $G_N\,$, plus
 \begin{equation}
  \Delta_\zk G \zH 2\zk T^{-1}_x G(t,x+y) =
  - 2\zk \frac1{(4\pi t)^{1/2}} e^{\zk x}
 \int_x^\zI e^{-\zk s}    \exp \left[-\,\frac{(s+y)^2}{4t}\right]  ds.
\label{RT}\end{equation}
 This expression can be rearranged into
 \[
 - \, \frac{\zk}{(\pi t)^{1/2}} e^{\zk (x+y)} e^{\zk^2 t}
 \int_x^\zI \exp \left [-\,\frac{(s+y+2\zk t)^2}{4t}\right ]  ds
=
 -\zk e^{\zk (x+y)} e^{\zk^2 t} 
 \erfc \left (\frac{x+y}{ \sqrt{4t}} + \zk \sqrt{t}\right ).
 \]
  Here $\,\erfc{}$ is the complementary error function,
 denoted by $1-\Phi$ in~\cite{GR}.
 Thus, finally, we have (for $\zk>0$)
 \begin{multline} 
 G_R(t,x,y) =
 \frac1{(4\pi t)^{1/2}} 
 \left \{ \exp \left [-\,\frac{(x-y)^2}{4t}\right ] 
 +\exp \left [-\,\frac{(x+y)^2}{4t}\right ]  \right \}
  \\
-\zk e^{\zk (x+y)} e^{\zk^2 t} 
 \erfc \left (\frac{x+y}{ \sqrt{4t}} + \zk \sqrt{t}\right ).
 \label{R}\end{multline}
 
 Formulas (\ref{RT}) and (\ref{R}) 
(which are not new \cite{Bryan1,CJ})
are the two key results of this 
subsection; (\ref{RT}) has a path-sum interpretation, which is best 
postponed to the next subsection.

 From (\ref{eig}), 
 the eigenfunction expansion of the Robin heat kernel is
 \begin{equation}
 G_R(t,x,y) = \frac2{\zp} \int_0^\zI
 \frac{\zw^2}{ \zw^2+\zk^2}
 \left [\cos(\zw x) + \frac\zk\zw\,\sin(\zw x) \right ]
\left [\cos(\zw y) + \frac\zk\zw\,\sin(\zw y) \right ]
 e^{-\zw^2 t}\, d\zw.
  \label{Heig}\end{equation}
 The integrals can be evaluated by formulas (3.954) of \cite{GR}
 (printed incorrectly in some earlier editions);
eventually the same result (\ref{R}) is obtained,
but the calculation via $T^{-1}$ is quicker.

 An integration by parts shows that
 \begin{equation}
 \int_0^\zI \zD_\zk G(t,x,\x,x,\x)\,dx =
 (4\zp t)^{-(d-1)/2} \,
 \tfrac12 \bigl( e^{\zk^2t} \erfc(\zk\sqrt t)-1\bigr). 
 \label{ht}\end{equation}
 Expanding (\ref{ht}) as a power series in $\zk$ or $t^{1/2}$
(see \cite[(8.253.1)]{GR}),
\begin{eqnarray}
 \tfrac12 \bigl( e^{\zk^2t} \erfc(\zk\sqrt t)-1\bigr) &=&         
 \frac12 \sum_{j=1}^\zI \frac{(\zk^2t)^j}{j!}
 -\frac1{\sqrt{\pi}} \sum_{j=0}^\zI 
\frac{2^j(\zk\sqrt{t})^{2j+1}}
{(2j+1)!!} \nonumber\\
&=&
 -\,\frac {\zk\sqrt{t}}{\sqrt{\pi}}+\frac{ \zk^2 t}2
-\frac{2\zk^3t^{3/2}}{3\sqrt{\pi}}
  +\frac{\zk^4t^2}4
 + \cdots,
\label{ankappa}\end{eqnarray}
 reproduces the known contributions \cite{BG,K5} of a 
 (flat, constant-$\zk$) Robin boundary to the 
usual heat-kernel trace expansion
and extends that information explicitly to all orders.
Perhaps more interesting is that (\ref{ht}) is a rare example of 
a heat trace known exactly for all~$t$, thereby leading to 
spectral densities known exactly for all~$\zw$;
we return to this point in Sec.~\ref{ssec:eigden}.

 \subsection{\label{ssec:quantum}The Schr\"odinger equation}

The Green function for the time-dependent Schr\"odinger equation,
 $i\pd ut = -\zV^2u$
 (in units where $\hbar=1$ and $m=\frac12$),
 also known as the quantum propagator,
  is obtained formally by replacing $t$ by $it$ in all the equations 
of the previous section.
 It can be seen that this takes the variable to the boundary of the 
domain of analyticity where all the integrals are meaningful.
 More interesting than the resulting formula in terms of the 
analytic continuation of $\,\erfc{}$ is the counterpart of the 
 prior formula (\ref{RT}),
 \begin{multline} 
 G_R(t,x,y) =
   \frac1{(4\pi it)^{1/2}} 
 \left \{ \exp \left [i\frac{(x-y)^2}{4t}\right ] 
 +\exp \left [i\frac{(x+y)^2}{4t}\right ]  \right \}
  \\
  - 2\zk \frac1{(4\pi it)^{1/2}} e^{\zk x}
 \int_x^\zI e^{-\zk s}    \exp \left[i\frac{(s+y)^2}{4t}\right]  
ds,
  \label{schrod}\end{multline}
 which admits an interpretation as a sum over paths.
 In analogy with the previous discussion of the wave equation,
 one would expect a source at 
$(0,y)$ to influence the solution at $(t,x)$ 
  along the direct path between the points
(the contribution of the free quantum kernel),
  and along the path that bounces 
elastically off the boundary (the image term of the Neumann 
solution), and possibly along paths that hit the boundary, stay 
there awhile,  and then return with the same energy;
 the problem is to show that the integral term  (\ref{RT}) lends 
itself to this last interpretation.
 The difference from the wave case is that in nonrelativistic 
 mechanics the paths may have any speed.
 Therefore, for any two points there always exist a direct path and 
an echo path, and also infinitely many paths of the delay type
(Fig~\ref{fig:paths}(b)).

 Recall first that for a freely moving particle the action of a 
trajectory segment of length $L$, speed $v$, and time $t$ is
 \begin{equation}
 S = \frac{L^2}{4t}=\frac{vL}4 = \frac{v^2 t}4 \,.
 \label{action}\end{equation}
 The action functional is additive over segments, as is clear from 
the last form given, in which $v^2/4$ is the constant (kinetic) 
energy of the orbit.
 So the quantities $(x\mp y)^2/4$ that appear in the exponents of 
the direct and echo terms of~(\ref{schrod})
  are the total actions of the corresponding paths.

 In the integral term in~(\ref{schrod}) we make the usual  change 
of integration variable $s= x+\ze$
and the further substitution
 \begin{equation}
\ze = vt -(x+y)
 \label{v}\end{equation}
  (thereby defining~$v$)
and also define
 \begin{equation}
 v_0 \zH \frac{x+y}t\,, \zJ u \zH t-\frac{x+y}v\,.
 \label{newvar1}\end{equation}
The term becomes
\begin{subequations} \label{qDel} \begin{eqnarray}
  \zD_\zk G &=&\frac{-2\zk}{ (4\zp it)^{1/2}}
  \int_{v_0}^\zI e^{-\zk t(v-v_0)} e^{iv^2t/4}\, dv
  \label{qDel1}\\ 
 &=&-\,\frac{2\zk t^{1/2}}{ (4\zp i)^{1/2}}
 \int_{v_0}^\zI e^{-\zk uv} e^{i[v(x+y) +v^2u]/4}
\, dv 
  \label{qDel2}
\end{eqnarray} \end{subequations}
 (where $u$ depends on $v$).
This equation has the following physical interpretation:
 $v$ is the speed of the ``particle'' as it travels from $y$ to the 
boundary and again from the boundary to $x$.
 Thus $v(x+y)/4$ is the action of those two trajectory segments,
 $(x+y)/v$ is the time consumed by them, and therefore $u$ is the 
remaining time, which the particle spends somehow attached to the 
wall. 
 The term $v^2u/4$, the product of this time with the energy of the 
orbit, is the action associated with this sojourn at the wall.
Each such orbit contributes to the propagator with an amplitude
 $-2\zk t e^{-\zk uv}$ times the usual amplitude,
  $(4\zp i t)^{-1/2}$.
 Also, $\ze/t = v-v_0$ is the difference (necessarily positive) 
 between the speed of this orbit and that of the echo orbit, a 
limiting case.

\emph{Remark:} The discovery of the action expression
  $A\zH v^2u/4$
 (and hence the correct relation between $\ze$ and $v$)
 was guided by the principle that the total action should be
 $v(x+y)/4 +A$, where $A$ depends only on the local physics at the 
boundary; that is, $A$ could be a function of $v$, $u$, and (in 
principle) $\zk$, but must not depend in any other way on $t$, $x$, 
and $y$. 
  That $A$ turns out to be precisely the total energy times 
the elapsed time $u$ was an unforeseen bonus, as was the fact that 
the exponential factor in the amplitude is likewise determined by 
the local physics.

 Since the integral (\ref{RT}) for the heat equation has the same 
structure, conceptually it also can be given a path-sum 
interpretation.
 Because of the diffusive nature of solutions of the heat equation, 
however, any trace of ``classical'' behavior is difficult to 
discern, even for sharply peaked initial data.

\subsection{\label{ssec:lap}Laplace's equation in one higher 
 dimension}

 The Green function that solves $\zV^2 u + \pd{u^2}{t^2}=0$
 (with $u$ bounded as $t\to+\zI$)
 for given boundary data $u(0,x,\x)$
 on the hypersurface $t=0$ is called the
Poisson kernel or cylinder kernel for the spatial geometry 
concerned.
 Its limiting behavior as $t\downarrow0$ can be used to determine 
the vacuum (Casimir) energy density of a scalar field in that 
geometry~\cite{systemat}.
When $d=1$ and the initial hypersurface is the whole line,
  the Poisson kernel is
 \begin{equation}
 G(t,x,y) = \frac1{\pi} \, \frac t{t^2+(x-y)^2}\,.  
 \label{P1}\end{equation}
 (Higher dimensions will be treated elsewhere~\cite{Fprog}.)

 In this problem the analogue of (\ref{RT}) is
\begin{equation}
 \zD G_\zk = -\, \frac{2\zk t}{\zp} \,e^{\zk x}
 \int_x^\zI \frac{ e^{-\zk s}}{ (s+y)^2 + t^2}\, ds.
 \label{cyl}\end{equation}
 Let $w=s+y$ and perform a partial-fraction decomposition:
 \[
 \zD G_\zk = -\, \frac{i\zk}{\pi} \, e^{\zk(x+y)} 
 \int_{x+y}^\zI e^{-\zk w} \left [\frac1{w+it} -\frac1{w-it}\right ] 
\,dw.
 \]
 By \cite[(3.352.2)]{GR}  one gets (see (\ref{Eidef}))
\begin{equation}
 \zD G_\zk = \frac{i\zk}{\pi} \, e^{\zk(x+y)} 
 \left [ e^{i\zk t} \Ei \bigl(-\zk(x+y+it)\bigr) 
 - e^{-i\zk t} \Ei \bigl(-\zk(x+y-it)\bigr)\right ] , 
 \label{Ei}\end{equation}
which can also be written
 \begin{equation}
   \zD G_\zk =\frac{2\zk}{\zp} \, e^{\zk(x+y)}\,
 \IM \left [e^{-i\zk t} \Ei \bigl(i\zk t- \zk (x+y)\bigr)\right ]
 \zj\text{(for $t$ real).} 
 \label{Ei2}\end{equation}
 To get the complete cylinder kernel for the Robin problem, add
 \begin{equation}
 G_N =\frac1{\pi} \left [ \frac t{t^2+(x-y)^2}
   + \frac t{t^2+(x+y)^2}\right ].
 \label{Ncyl}\end{equation}

 The eigenfunction expansion analogous to (\ref{Heig}) 
 just has $e^{-\zw t}$ in place of $e^{-\zw^2t}$.
 From there a lengthy calculation using 
 \cite[(3.354.1,2) and (8.233.1)]{GR} 
verifies that $G_R - G_N= \zD G_\zk $ as given in~(\ref{Ei}).
In summary,  this problem is very similar to the heat problem 
 in Sec.~\ref{ssec:heat}, 
 but with different special functions appearing.

The cylinder kernels discussed here solve Laplace's equation with 
nonhomogeneous Dirichlet data at $t=0$; they suffice for 
calculating total Casimir energy and for calculating vacuum energy 
density when the latter is defined with the value of the 
``conformal coupling constant'' set to $\xi = \frac14$
 \cite{systemat,funorman}.
 Obtaining the energy density for other values of $\xi$ requires 
the Green function for the problem with Neumann ``initial'' data,
 $\pd ut(0,x,\x)$.
 For the whole real line, that kernel is
 \begin{equation}
 G  = \frac1{2\zp} \,\ln [ t^2 + (x-y)^2].
 \label{logcyl}\end{equation}
Again, Theorem~\ref{th:green} can be implemented exactly to solve the 
 temporal Neumann problem with the Robin condition at the spatial 
boundary, $x=0$;
  the relevant integral is \cite[(4.337.1)]{GR}, and the result is
 \begin{equation}
 G_R =  \frac1{2\zp} \{\ln [ t^2 + (x-y)^2] -\ln [ t^2 + (x-y)^2] \}
+\frac2{\pi} \, e^{\zk(x+y)}\,\RE \left [e^{-i\zk t} \Ei\bigl(i\zk t -
\zk(x+y)\bigr) \right ].
 \label{logcylR}\end{equation}
 (Note that the first term is $G_D\,$, the kernel for the 
 \emph{Dirichlet} homogeneous spatial boundary condition;
  the reflection term 
 in $G_N$ has been overwhelmed by an identical term in $\zD_\zk G$
 with a factor $-2$.)

\emph{Remark:}
  In keeping with the well known ``Green's identity'' structure of 
the solution formulas for second-order elliptic boundary-value 
problems, the cylinder kernel 
 for the nonhomogeneous  temporal Dirichlet condition is the 
$t$~derivative of the corresponding cylinder kernel 
 for the temporal Neumann condition. 
 This relationship is easily checked for both pairs,
 (\ref{P1})/(\ref{logcyl}) and
 (\ref{Ei2})/(\ref{logcylR}).

 Just as the Schr\"odinger kernel is an analytic continuation of 
the heat kernel, replacing $t$ by $it$ in a cylinder kernel leads 
to a certain fundamental solution of the wave equation.
 More precisely, in (\ref{P1}) and (\ref{Ei}) (where $t$ was positive)
  one 
should replace $t$ by $i(t-i0)$, where now $t$ can have either 
sign but the negative infinitesimal imaginary part is needed 
because a singularity is encountered on the real axis when 
 $|t|> x+y$.
 For the basic Green function in (\ref{P1}) the singularity is a pole, 
which has the well known decomposition
 \begin{equation}
 \frac{i}{\pi}\, \frac{t}{ (x-y)^2 - t^2} =
 \frac12\,[\delta(x-y-t)+\delta(x-y+t)]
 +\frac i{2\pi} \, \mathcal{P}
  \left [\frac1{x-y-t} -\frac1{x-y+t}\right ].
  \label{Wight}\end{equation}
 The delta term is recognized as the d'Alembert Green function 
 (\ref{dalem}),
  which solves the wave equation with given initial value 
and vanishing initial time derivative.
 The principal-value term appears because the asymptotic condition
 on the Poisson kernel at $t=+\zI$ has evolved into a
 positive-frequency condition on the solution of the wave equation,
 hence an initial value of $\pd ut$ that is a certain nonlocal 
functional of the initial value of~$u$.
 In the language of quantum field theory, (\ref{Wight}) 
 is (proportional to)
 the time derivative of the Wightman function, and the d'Alembert 
term is proportional to the time derivative of the field 
commutator.
 (Starting from the other kind of cylinder kernel would avoid the time 
derivatives.)

 All these statements have analogues for the Green functions of the 
Robin problem.
 From (\ref{Ei}) we get as the correction to the Neumann Wightman 
function
 \begin{equation}
 \zD G_\zk = \frac{i\zk}{\pi}\, e^{\zk(x+y)}
 [e^{-\zk t} \Ei \bigl(\zk(t-x-y)\bigr)
 - e^{\zk t} \Ei \bigl(\zk(t+x+y)\bigr)].
\label{deltaWight}\end{equation}
 Here $\,\Ei\,$ is defined by \cite[(8.211.1)]{GR}:
 \begin{equation}
 \Ei(z) = \int_{-\zI}^z \frac{e^t}{t}\,dt
 \label{Eidef}\end{equation}
 with a branch cut on the positive real axis (which comes into play 
when        $|t|> x+y$).
 According to \cite[(8.240.3)]{GR}, 
 $
 \Ei (x\mp i0) = \mathcal{P}\Ei(x) \pm i\pi.
 $
 (See also \cite[(3.352.5)]{GR} and the footnote 
on p.~228 of \cite{AS}.)
 Thus (\ref{deltaWight}) consists of a principal-value term,
  plus a jump term that equals
 $
 -\zk e^{\zk(x+y-t)} \zy(t-x-y)
 $
 in the case of positive~$t$ (for instance).
 This last is recognized as the $\zD_\zk G$  found for the 
classical wave  propagator in~(\ref{GR}).

As in the quantum problem one can in  these problems introduce 
the variables 
 \begin{equation}
 v = \frac{x+y+\ze}{t}\,,\zj u = t- \frac{x+y}{ v}\,,
 \label{newvar2}\end{equation}
 and interpret them as the speed of a reflected path from $y$ 
to~$x$ and the time delay of the path at the reflection point.
 All that changes is the way in which the part of the integrand 
coming from the original Green function depends on~$v$;
 in the case (\ref{cyl}) one has 
 \begin{equation}
 \zD_\zk G= -\,\frac{2\zk}{\pi} 
 \int_{v_0}^\zI e^{-\zk t(v-v_0)}
 \frac{dv}{ v^2 +1}
 \label{Wightpath}\end{equation}
 in place of~(\ref{qDel1}).
 As $v$ varies from the kinematic minimum, $v_0= (x+y)/t$, to 
infinity,
 $u$ varies from $0$ to the kinematically allowed maximum,~$t$.
 These reparametrizations are not useful calculationally in the 
simple problems treated in this paper, but they are likely to 
become important in arriving at a physically correct approximation 
ansatz in more complicated problems where the relation between time 
displacement ($u$) and space displacement ($\ze$) is nonlocal.
 The interaction between the field and the Robin boundary should 
take place at the boundary, not throughout a spatial layer of size 
 $\ze\approx \zk^{-1}$.

 \section{\label{sec:adv}Strategies for advanced applications}

 One would like to extend the Dirichlet-to-Robin technique to 
problems where the boundary is curved, $\zk$~is not constant, or 
$\zV^2-\zd$ is replaced by a differential operator whose coefficients 
depend on~$x$.
 There are two obstacles to be overcome.
 First, one must have a valid Dirichlet solution, $v$, or Neumann 
Green function, $G_N\,$, from which to start, and in general even 
those elementary boundary conditions cannot be solved exactly by 
the method of images. 
 That problem leads into the general subject of semiclassical (or 
other) approximations and will not be further considered here.
 Second, in general $T$ will not commute with the differential 
operator, and hence $u=T^{-1} v$ will not satisfy the same 
partial differential equation as $v$.
 There are two strategies one might pursue to get around this 
problem.

 The first (which is not our favorite) is to find a differential 
equation to be satisfied by the Dirichlet function~$v$
 that will cause the Robin function $u$ to satisfy the correct 
differential equation.
 Note that there is some freedom in how to extend the definition 
 (\ref{Toper}) of 
$T$ to the interior region, since $\zk$ is given only on the 
boundary.
 For example, consider the wave equation on the interval $0<x<1$ 
with a second Robin condition at the right end, $-\pd ux(0,1) = 
\zk'u(0,1)$ (with negative sign because the normal derivative now 
points in the opposite direction).
 If $v(t,x)$ solves the (doubly) Dirichlet problem on the interval,
 then $T^{-1}v$ obeys the correct Robin condition at $x=0$ but 
not at $x=1$, unless $\zk'=-\zk$.
 Not surprisingly, $\zk'=-\zk$ is a condition for the eigenvalues 
  of the Robin problem to be the same as those of the Dirichlet 
  problem (see \cite[(4.4)]{RS});
 it is clear that the construction $G_R=T^{-1}G_DT$ is not possible 
for any $T$ unless the problems are isospectral.
 It is possible, however, to choose a function $\zk(x)$ that 
smoothly interpolates between $\zk(0)=\zk$ and $\zk(1)= -\zk'$
 and to define $T$  accordingly so that  Lemma~\ref{lem:taut}
  holds at both endpoints.  
 The solution for $T^{-1}$ now is more complicated than 
 (\ref{T1}) or (\ref{1DT2}) but still elementary.
Finally,  the equation that $v=Tu$ must obey to cause 
$u$ to obey the wave equation is
 \begin{eqnarray} \pd {^2v}{t^2} -\pd {^2v}{x^2} 
&=& 2 \od {\zk}x \, \pd ux + \od{^2u}{x^2} \, u 
 \nonumber\\
 &=& 2 \od {\zk}x \, v + \od{}x\left (\od{\zk}x + \zk^2\right )u, 
\label{deformeq} \end{eqnarray}
 an integrodifferential equation in general.
 With luck one can choose $\zk(x)$ so that the coefficient of $u$ 
vanishes, and then one has a slightly modified wave equation 
for~$v$ 
 (with a spatial operator whose Dirichlet realization \emph{is} 
isospectral to our Robin problem).
 For some values of $\zk(0)$ and $\zk(1)$, however, the
$\zk(x)$ satisfying the required conditions has a pole inside the 
interval $(0,1)$, so that this construction fails.
 
 In higher dimensions such difficulty appears to be generic.
 Defining $T$ throughout a disk, for example, requires 
extrapolating the normal vector to the boundary smoothly throughout 
the interior.  Such a vector field must vanish somewhere, and at 
that point the first-order partial differential equation to be 
solved to construct $T^{-1}$ becomes singular.

 We conclude that although this approach may allow some special 
problems to be solved by tricks, it is not promising as a 
systematic method.

 The second strategy is to exploit the locality of the boundary 
interaction and the classical-path structure of the field dynamics.
 Our expectation is that a locally defined Dirichlet-to-Robin 
transform $T^{-1}$ tells how any solution locally reflects off a 
Robin boundary, and that this information can be combined with 
standard semiclassical technology in the bulk (and for Dirichlet 
and Neumann reflections from curved boundaries) to construct global 
approximate solutions.
 In the next section we implement this approach for what is 
probably the simplest situation, two parallel flat boundaries with 
empty Euclidean space between them.

 \section{\label{sec:interval}An intermediate application:  
 Wave equation and spectrum on an interval}

  \subsection{\label{ssec:prob}The problem} 

 Consider first the one-dimensional wave equation on an interval
with a Dirichlet boundary at the right end
 and a Robin boundary at the left:
\begin{subequations} \label{w}
 \begin{equation}
\pd{^2u}{t^2} = \pd{^2u}{x^2} \zj (0<x<L, \ 0<t<\zI), 
 \label{waveeqint}\end{equation}
 \begin{equation}
 u(0,x) = f(x), \zJ \pd ut(0,x) = 0, 
 \label{initdata}\end{equation}
 \begin{equation}
\pd ux(t,0) = \zk u(t,0) \zj (\zk\ge0), \zJ u(t,L) = 0.
 \label{2bc}\end{equation}
 \end{subequations}
 (There would be no difficulty in principle in handling a second Robin 
condition at $x=L$.)
 The makeup of the solutions from right- and left-moving pulses 
makes this model particularly easy and instructive.
 As shown in Sec.~\ref{sec:adv}, it is not possible to obtain a solution
 by applying $T^{-1}$ once and for all to the solution of the pure 
Dirichlet problem 
 (although the most severe complications mentioned in Sec.~\ref{sec:adv}
  are not present here).
 Instead, the transform will be applied repeatedly as each pulse 
 strikes the Robin boundary.

 \subsection{\label{ssec:neum}The Neumann--Dirichlet case}

\begin{figure}
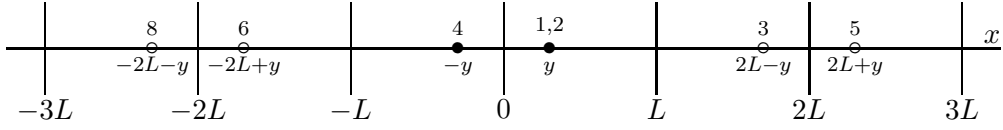
 
 $$\beginpicture
\setcoordinatesystem units <0.8truein,0.8truein> point at 0 0
 \putrule from -3.25 0 to 3.25 0
 \put{$x$} [br] <0pt, 2pt> at 3.25 0
 \putrule from 0 -0.3 to 0 0.3
 \putrule from 1 -0.3 to 1 0.3
 \putrule from 2 -0.3 to 2 0.3
 \putrule from 3 -0.3 to 3 0.3
 \putrule from -1 -0.3 to -1 0.3
 \putrule from -2 -0.3 to -2 0.3
 \putrule from -3 -0.3 to -3 0.3
\put{$0$} [t] <0pt,-2pt> at 0 -0.3
\put{$L$} [t] <0pt,-2pt> at 1 -0.3
\put{$2L$} [t] <0pt,-2pt> at 2 -0.3
\put{$3L$} [t] <0pt,-2pt> at 3 -0.3
\put{$-L$} [t] <0pt,-2pt> at -1 -0.3
\put{$-2L$} [t] <0pt,-2pt> at -2 -0.3
\put{$-3L$} [t] <0pt,-2pt> at -3 -0.3
\put{$\bullet$} at 0.3 0
 \put{$\circ$} at 2.3 0
\put{$\circ$} at -1.7 0
\put{$\bullet$} at -0.3 0
\put{$\circ$} at -2.3 0
 \put{$\circ$} at 1.7 0
\put{$\s y\vphantom{L}$} [t] <0pt, -3pt> at 0.3 0
 \put{$\s 1,2$} [b] <0pt, 5pt> at 0.3 0
\put{$\s 2L+y$} [t] <0pt, -3pt> at 2.3 0
 \put{$\s 5$} [b] <0pt, 5pt> at 2.3 0
\put{$\s -2L+y$} [t] <0pt, -3pt> at -1.7 0
 \put{$\s 6$} [b] <0pt, 5pt> at -1.7 0
\put{$\s -y\vphantom{L}$} [t] <0pt, -3pt> at -0.3 0
 \put{$\s 4$} [b] <0pt, 5pt> at -0.3 0
\put{$\s -2L-y$} [t] <0pt, -3pt> at -2.3 0
 \put{$\s 8$} [b] <0pt, 5pt> at -2.3 0
\put{$\s 2L-y$} [t] <0pt, -3pt> at 1.7 0
 \put{$\s 3$} [b] <0pt, 5pt> at 1.7 0
 \endpicture$$
 \caption{Image sources for the wave equation on interval $(0,L)$
 with Neumann boundary at~$0$ and Dirichlet boundary at~$L$. 
 Filled circles indicate positive terms, open circles indicate 
negative terms.
The numerical labels are for comparison with 
Fig.~\ref{fig:wavepath}.}
\label{fig:images} \end{figure}

 To start, recall what happens when $\zk=0$.
 The method of images associates the generic point $y$ in the 
interval with an infinite set of image points in the unphysical 
regions, as shown in Fig.~\ref{fig:images},
  where the open circles indicate the 
images that are weighted negatively.
 By d'Alembert's formula (\ref{dalem}), therefore, 
 the wave Green function is
 \begin{multline}
 G_N(t,x,y) = \frac12 \sum_{n=-\zI}^\zI (-1)^n \bigl[
 \zd(t+x-y-2nL) +\zd(t-x+y-2nL) 
 \\
 -\zd(t+x+y-2(n+1)L) -\zd(t-x-y-2(n-1)L)\bigr].
 \label{GNint}\end{multline}
 With this choice of indexing,
 the pulses relevant for $t>0$ are those with $n\ge 0$ in the first 
three terms and those with $n\ge1$ in the fourth term, and
 when we go on to the Robin problem, $n$~will be the number of 
times the operator $T^{-1}_\zk$ needs to be applied.

 The \emph{trace} of the wave kernel is the Fourier cosine 
transform (with respect to $\zw$) of the eigenvalue density.
 Here it is
 \begin{multline*}
 \int_0^L G_N(t,x,x)\,dx = 
 \sum_{n=-\zI}^\zI (-1)^n \bigl[
 L \zd(t-2nL) 
 \\
  - \frac14 \zy\bigl(0<-{\tfrac t2} +(n+1)L <L\bigr)
 -\frac14  \zy\bigl(0<{\tfrac t2} -(n-1)L <L\bigr)\bigr], 
\end{multline*}
 where $\zy(a<t<b) = \zy(t-a)\zy(b-t)$ is the characteristic 
function of interval $(a,b)$.
 The inequality in the last term is equivalent to
 $2(n-1)L < t < 2nL$, while that in the next-to-last term is
 equivalent to $2nL< t<2(n+1)L$.
 Reindexing then shows that these two terms cancel.
 (See also (\ref{nlocal}), however.)
   Thus one has
 \begin{equation}
\int_0^L G_N(t,x,x)\,dx = L \sum_{n=-\zI}^\zI (-1)^n\zd(t-2nL),
  \label{GNinttrace}\end{equation}
  a sum over all the periodic orbits (with lengths $2|n|L$).
 Its inverse Fourier cosine transform is
 \begin{eqnarray}
  \zr_N(\zw) &=& \frac2{\zp} \int_0^\zI dt\, \cos(t\zw) 
 \int_0^L G_N(t,x,x)\,dx \nonumber
  \\
 &=& \frac{2L}{\zp} \sum_{n=0}^\zI (-1)^n
  \left (1-\frac12\delta_{n0}\right )
 \cos(2nL\zw). 
 \label{rhon}\end{eqnarray}
This sum can be evaluated by the Poisson summation formula 
\cite{poisref} as
 \begin{eqnarray}
  \frac L\pi \sum_{n=-\zI}^\zI e^{in\zp} e^{2inL\zw}
 &=& \frac L\pi \sum_{k=-\zI}^\zI \int_{-\zI}^\zI 
 e^{-2\pi ikn} e^{i\zp n} e^{2inL\zw} \, dn \nonumber
 \\
 &=& \frac L\pi \sum_{k=-\zI}^\zI 2\pi  \zd(-2\zp k +\zp +2L\zw) 
 \nonumber \\
 &=& \sum_{k=1}^\zI
  \delta\left (\zw-\frac{\pi}{L}\bigl(k-{\tfrac12}\bigr)\right ) 
\label{neupois}\end{eqnarray}
 (since only $\zw\ge 0$ is meaningful in the cosine transform).
 That is, the eigenvalues (or, rather, their square roots, the 
eigenfrequencies) are 
 $\frac{\pi}{2L}$, $\frac{3\pi}{2L}$, $\ldots\,$, as expected.

\emph{Remark:} The cancellation of the contributions from 
``bounce'' orbits (closed but not periodic) is an artifact of the 
mixed Neumann--Dirichlet boundary conditions.
 In general those orbits produce the ``surface area'' term in the 
Weyl expansion of the eigenvalue density (cf.~\cite{SSCL}).
 For example, in the \emph{pure} Neumann case the $(-1)^n$ would be 
missing from the formulas above, so that
 \begin{eqnarray}
 \int_0^L G(t,x,x)\,dx &=& L \sum_{n=-\zI}^\zI \zd(t-2nL)
 +\frac12\sum_{n=-\zI}^\zI \zy\bigl(2nL<t<2(n+1)L\bigr) 
 \nonumber\\
&=&  L \sum_{n=-\zI}^\zI \zd(t-2nL) +\frac12\,. 
 \label{puretr}\end{eqnarray}
 In the pure Dirichlet case the $\frac12$ becomes $-\frac12$.
 In the inverse cosine transform the $\pm \frac12$ produces
 $\pm \frac12\delta(\zw)$ (cf.\ Appendix~\ref{app:delta}),
  which combines with a 
 $+ \frac12\zd(\zw)$ in the Poisson sum to denote the presence or 
absence, respectively, of the eigenvalue at $\zw=0$.

 \subsection{\label{ssec:wk}The wave kernel}

Turn now to the Robin case.  The wave kernel $G_R$ will have the 
same basic pulse structure as $G_N\,$,  but every time a pulse 
reflects from the left boundary it will acquire a time-delayed 
component; these effects cumulate as indicated in 
Fig.~\ref{fig:wavepath}.
At the first step we need to  know how to produce the pulse 
numbered 4 in the figure by Robin reflection of pulse~1.
 Since pulse~1 by itself is not a Green function, 
 Lemma~\ref{lem:Tgreen} and Theorem~\ref{th:green}
 do not apply directly, but they carry over in essence:

 \begin{figure}
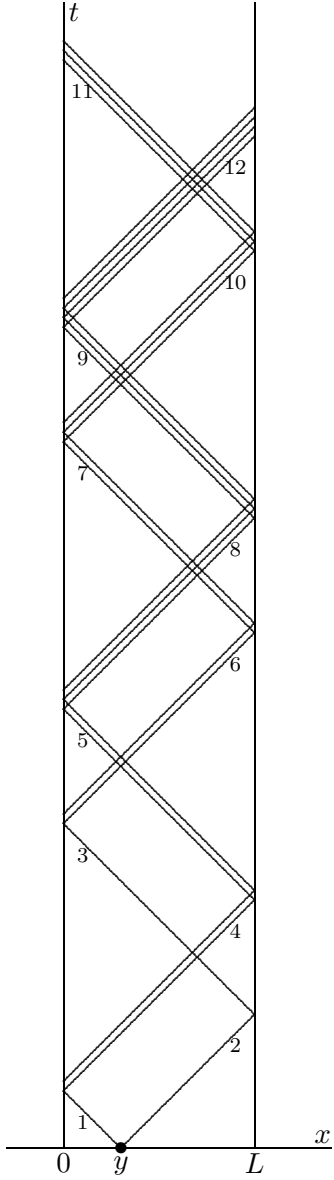
 
 $$\beginpicture
  \setcoordinatesystem units <1truein,1truein>  point at 0 0
 \putrule from 0 0 to 0 6
 \putrule from 1 0 to 1 6
 \putrule from -0.3 0 to 1.4 0
 \put{$x$} [rb] <0pt, 2pt> at 1.4 0
 \put{$t$} [lt] <2pt,0pt> at 0 6
 \put{$y$} [t] <0pt,-3pt> at 0.3 0
 \put{$0$} [t] <0pt,-2pt> at 0 0
 \put{$L$} [t] <0pt,-2pt> at 1 0
 \put{$\bullet$} at 0.3 0
 \plot 0.3 0
        0 0.3 /
 \put{$\s1$} [t] <0pt,-2pt> at 0.1 0.2
 \plot 0.3 0 
        1 0.7 /
  \put{$\s2$} [t] <0pt,-2pt> at 0.9 0.6
 \plot 1 0.7 
        0 1.7 /
  \put{$\s3$} [t] <0pt,-2pt> at 0.1 1.6
 \plot 0 0.3
        1 1.3 /
 \plot 0 0.35
        1 1.35 /
   \put{$\s4$} [t] <0pt,-2pt> at 0.9 1.2
 \plot 1 1.3
        0 2.3 /
 \plot 1 1.35
        0 2.35 /
   \put{$\s5$} [t] <0pt,-2pt> at 0.1 2.2
 \plot 0 1.7
        1 2.7 /
 \plot 0 1.75
        1 2.75 /
    \put{$\s6$} [t] <0pt,-2pt> at 0.9 2.6
 \plot 1 2.7
        0 3.7 /
 \plot 1 2.75
        0 3.75 /
   \put{$\s7$} [t] <0pt,-2pt> at 0.1 3.6
 \plot 0 2.3
        1 3.3 /
 \plot 0 2.35
        1 3.35 /
 \plot 0 2.4
        1 3.4 /
    \put{$\s8$} [t] <0pt,-2pt> at 0.9 3.2
 \plot 1 3.3
        0 4.3 /
 \plot 1 3.35
        0 4.35 /
 \plot 1 3.4
        0 4.4 /
   \put{$\s9$} [t] <0pt,-2pt> at 0.1 4.2
 \plot 0 3.7
        1 4.7 /
 \plot 0 3.75
        1 4.75 /
 \plot 0 3.8
        1 4.8 /
    \put{$\s10$} [t] <0pt,-2pt> at 0.9 4.6
 \plot 1 4.7
        0 5.7 /
 \plot 1 4.75
        0 5.75 /
 \plot 1 4.8
        0 5.8 /
   \put{$\s11$} [t] <0pt,-2pt> at 0.1 5.6
 \plot 0 4.3
        1 5.3 /
 \plot 0 4.35
        1 5.35 /
 \plot 0 4.4
        1 5.4 /
 \plot 0 4.45
        1 5.45 /
    \put{$\s12$} [t] <0pt,-2pt> at 0.9 5.2
 \endpicture$$
  \caption{Structure of the solution (\ref{W}) of the wave equation 
  on interval $(0,L)$ 
 with Robin boundary at~$0$ and Dirichlet boundary at~$L$.}
\label{fig:wavepath} \end{figure}

 \begin{lemma}\label{lem:reflect}
Let $u(t,x)$ be any solution of the one-dimensional wave equation.
 Then
 \begin{eqnarray*}
  u_R(t,x) &=& u(t,x) + u(t,-x) +2\zk T^{-1}_x [u(t,-x)] 
 \\
 &=& u(t,x) + u(t,-x) -2\zk \int_x^\zI e^{\zk(x-s)} u(t,-s)\, ds 
\end{eqnarray*}
 is a solution satisfying the Robin condition.
 \end{lemma}

\emph{Proof:}  $u(t,-x)$ satisfies the wave equation by virtue of 
the latter's reflection symmetry.  Then the third term is also a 
solution by Lemma~\ref{lem:tback}.
 It remains to check the boundary condition:
 \begin{eqnarray*}
  Tu_R(t,x) &=&
 Tu(t,x) + T_x[u(t,-x)] +2\zk u(t,-x)
  \\
 &=&\pd ux(t,x) -\zk u(t,x) -\pd ux(t,-x) +\zk u(t,-x), 
 \end{eqnarray*}
hence $Tu_R(t,0) = 0$.

 \emph{Remarks:} (1) The lemma is to be used in cases where 
$u$ represents a wave  impinging on the Robin 
boundary.
 Then the rest of $u_R$ represents a wave reflected in the opposite 
direction.  The reflection is ``causal'' in the sense that the 
incident wave is not modified by the construction.
 (2) This construction in its simple form may break down in more 
general problems (e.g., Schr\"odinger's equation with a potential), 
because 
 (a) $u(t,-x)$ may not satisfy the differential equation, or even 
be defined; (b) $T^{-1}$ may not commute with the differential 
operator.  See remarks in Sec.~\ref{sec:adv}.

 Pulse 1 is simply the term $\frac12\delta(t+x-y)$ in $G_N\,$.
 Apply Lemma~\ref{lem:reflect} to get pulse~4:
\[
 \frac12 \delta(t-x-y) - 2\zk\int_x^\zI e^{\zk(x-s)}
  \delta(t -s-y)\, ds 
 = \frac12 \delta(t-x-y) - 2\zk e^{-\zk (t-x-y)} \zy(t-x-y).
 \]
 Continuing inductively, one builds up the entire wave kernel as 
presented in the following theorem.

 \begin{theorem}\label{th:W}
For $t\ge0$ in the wave problem (\ref{w}) the Green function is
 \begin{subequations}\label{W}
 \begin{eqnarray} 
 G_R(t,x,y) &=&
\sum_{n=0}^\zI (-1)^n \Bigl[ \tfrac12\zd(t+x-y-2nL)  
 \nonumber\\
 && {}-\zk L^1_{n-1}\bigl(2\zk(t+x-y-2nL)\bigr)
 e^{-\zk(t+x-y-2nL)} \zy(t+x-y-2nL)\Bigr] 
 \label{W1}\\
 &+& \sum_{n=0}^\zI (-1)^n \Bigl[ \tfrac12\zd(t-x+y-2nL)  
 \nonumber\\
 &&{} -\zk L^1_{n-1}\bigl(2\zk(t-x+y-2nL)\bigr)
 e^{-\zk(t-x+y-2nL)} \zy(t-x+y-2nL)\Bigr] 
 \label{W2}\\
 &+&\sum_{n=0}^\zI (-1)^{n+1}
  \Bigl[ \tfrac12 \zd(t+x+y-2(n+1)L)  
 \nonumber\\
 &&{} -\zk L^1_{n-1}\bigl(2\zk(t+x+y-2(n+1)L)\bigr)
 \nonumber\\
&&\zJ\zJ{}\zM e^{-\zk(t+x+-y-2(n+1)L)} 
 \zy\bigl(t+x+y-2(n+1)L\bigr)\Bigr]
 \label{W3} \\
 &+&\sum_{n=1}^\zI (-1)^{n-1}
  \Bigl[ \tfrac12\zd(t-x-y-2(n-1)L) 
 \nonumber \\
 &&{} -\zk L^1_{n-1}\bigl(2\zk(t-x-y-2(n-1)L)\bigr)
 \nonumber\\
&&\zJ\zJ{}\zM  e^{-\zk(t-x-y-2(n-1)L)}
  \zy\bigl(t-x-y-2(n-1)L\bigr)\Bigr] 
  \label{W4}\end{eqnarray}
 \end{subequations}
Here $L^1_{n-1}$ is the Laguerre polynomial \cite[(8.970.1) and 
(8.971.2)]{GR}
 \begin{equation} 
L^1_{n-1}(x) = - L^{0\prime}_n(x) =
 \sum_{j=1}^n \binom nj \frac{(-x)^{j-1}}{(j-1)!}
 \zJ(L^1_{-1}\zH 0) \,.
 \label{laguerre}\end{equation}
 \end{theorem}

\emph{Remark:}
  In terms of the labeling in Fig.~\ref{fig:wavepath},
 term (\ref{W1})
 comprises pulses $4n+1$, (\ref{W2}) pulses $4n+2$,
 (\ref{W3})   pulses $4n+3$, and (\ref{W4})  pulses $4n$.
 (Note that $n$ starts from $1$ in~(\ref{W4}), but from $0$ in 
the other three terms.)

 The proof is a straightforward induction guided by 
Fig.~\ref{fig:wavepath}.
 Robin reflection  (by Lemma~\ref{lem:reflect}) of pulses $4n+1$ and $4n+3$ 
 produces pulses $4(n+1)$ and $4(n+1)+2$, respectively.
Similarly,  Dirichlet reflection at $x=L$ of pulses $4n$ and $4n+2$ 
yields  pulses $4n+1$ and $4n+3$.

 \subsection{\label{ssec:trace}The wave trace}

 Let us now take the trace of (\ref{W}).
 The terms  (\ref{W1}) and (\ref{W2}) give identical contributions,
  totalling
 \begin{equation}
  \int_0^L G_R(t,x,x)_\text{per}\,dx =  N + P, 
\label{pertrace}\end{equation}
 where
 \begin{equation}
N\zH   L \sum_{n=0}^\zI (-1)^n\zd(t-2nL)
 \label{neutrace}\end{equation}
 coincides (for $t\ge0$) with the result (\ref{GNinttrace}) 
 found previously for the Neumann boundary condition  and
 \begin{equation}
P\zH
 -2\zk L \sum_{n=1}^\zI (-1)^n
  L^1_{n-1}\bigl( 2\zk (t-2nL)\bigr)e^{-\zk(t-2nL)} \zy(t-2nL)
 \label{P}\end{equation}
 is the additional contribution  in the 
Robin problem of the periodic orbits.
  Note that both these terms carry an overall 
factor~$L$ --- they came from an integrand independent of~$x$ ---
 which is lacked by the remaining terms, which come from the bounce 
orbits and reflect the latter's close association with the 
boundaries rather than the global geometry.
 The contribution of the bounce orbits simplifies to
 \begin{equation}
 \int_0^L G_R(t,x,x)_\text{bou}\,dx =  A + B,
 \label{GRbou}\end{equation}
 where
 \begin{equation}
A\zH \tfrac12 (e^{-\zk t} -1)
\label{A} \end{equation} 
 and
 \begin{equation}
B\zH \sum_{n=1}^\zI \sum_{m=1}^n c(n,m) \frac{\zk^m}{m!}
 (t-2nL)^m e^{-\zk(t-2nL)} \zy(t-2nL),
\label{B}\end{equation}
 \begin{equation}
c(n,m)\zH \sum_{j=m}^n (-1)^{n-j} 2^{j-1} \frac{2n-j}{n}
\binom nj.
 \label{c}\end{equation}

\emph{Remarks:} (1) For fixed $t$ all sums encountered so far are 
finite, so there is no issue of convergence or term ordering.
 This property will be lost at the next step!
(2) As expected, all terms except $N$ vanish as $\zk\to0$.
 As $\zk\to+\zI$, it can be shown that $B\to0$, $A\to -\frac12$,
 and 
 \[P\to L\sum_{n=1}^\zI \bigl(1-(-1)^n\bigr) \zd(t-2nL);\]
 thus
 \begin{equation}
 \int_0^L G_R(t,x,x)\,dx \to L\sum_{n=0}^\zI \zd(t-2nL) - 
\tfrac12 \zj\text{as $\zk\to\zI$},
 \label{dirilim}\end{equation}
 which, as previously remarked, is the correct formula for the 
Dirichlet problem.
 (3) The critical dimensionless parameter of this model is 
$L\zk$.
 So far all the calculations are exact, so it has not been 
necessary to assume $L\zk$ either large or small.
 (4) We omit the lengthy calculation leading to ($B$), except to 
mention that it involves definite integrations over intervals of 
the form $t-2(n+1)L < s < t-2nL$, after which the index $n$ in the 
contributions from lower limits of integration need to be shifted 
relative to those from upper limits to combine terms.

 \subsection{\label{ssec:eigden}The eigenvalue density}

 We have already  calculated $\zr_N$ (\ref{rhon}), 
 the inverse cosine transform of $N$.
 That of $A$ is 
 \begin{equation}
\zr_{\zk,\text{av}}(\zw) = \frac1{\pi} \, \frac{\zk}{\zw^2+\zk^2}
  - \tfrac12\zd(\zw)\zJ (\zk\ne 0) .
 \label{rhoav}\end{equation}
 (Contrary to appearance, this object does approach the Neumann 
limit of $0$ 
 as $\zk\to0$, because the first term  converges to 
$\frac12\zd(\zw)$ in the distributional sense.
 The corresponding spectral-staircase formula,
 \begin{equation}
N_{\zk,\text{av}}(\zw) =\frac1\pi \tan^{-1} \frac{\zw}{\zk} 
 -\frac12 \zj\text{for $\zw >0$},
 \label{nav}\end{equation}
  looks less anomalous.) 
  Because $A$ is recognized as the wave trace (\ref{wt}) associated with 
the Robin boundary sitting in infinite space,
 one can interpret $\zr_{\zk,\text{av}}$ as the contribution of a 
 Robin boundary (relative to the Neumann base case) to the 
``averaged'' or ``smoothed'' spectral density  in dimension~$1$,
 analogous to
 \begin{equation}
\zr_{\zk,\text{av}}(\zw) = 
 \frac{\text{perimeter}}{2\pi} 
 \left (\frac{\zw}{\sqrt{\zw^2+\zk^2}}-1\right )
\label{av2D} \end{equation}
 in dimension $2$ (from \cite[(7)]{SPSUS}) and
 \begin{equation}
\zr_{\zk,\text{av}}(\zw) = \frac{\text{surface area}}{2\pi^2/\zw} 
\left [\tan^{-1} \left (\frac{\zw}{\zk}\right ) - \frac{\pi}{2}\right ]
\label{av3D}\end{equation}
 in dimension $3$ (from \cite[equation in abstract]{BB1}).
 These formulas are strikingly dimension-dependent, yet
 it must be possible to obtain them all as inverse Laplace 
transforms of the heat trace~(\ref{ht}).
 (Indeed, (\ref{rhoav})--(\ref{av3D}) can be verified  with the aid of 
the Laplace-transform table in~\cite{AS}.)
 The culprit, obviously, is the $d$-dependent power of~$t$ 
 in~(\ref{ht}).
When (\ref{ht}) is expanded as a power series
(see (\ref{ankappa})), as is traditionally done,
 the inverse Laplace transforms become elementary and yield 
series for $\zr_{\zk,\text{av}}$ in powers of $\zw^{-1}$
 whose coefficients depend on~$d$ in a relatively simple way.
 
\emph{Remark:} 
In general, $\rho_{\zk,\text{av}}$ for $d+2$ is essentially 
the 
antiderivative of $\rho_{\zk,\text{av}}$ for $d$, as demonstrated 
by (\ref{av3D}) and (\ref{nav}). 
This fact can be seen either as a result of multiplying the 
Laplace transform by $t^{-1}$, or as a result of convolving the 
eigenvalue density, with respect to~$\zw^2$, with that of 
Euclidean $\mathbf{R}^2$, which is constant.

 To treat $P$ and $B$ we need the integral formula 
 \cite[(3.944.5,6)]{GR} 
 \begin{equation}
\int_0^\zI  \zt^{j-1} e^{-\zk\zt} \cos(\zw\zt+\zd)\, d\zt =
 (j-1)!\, (\zw^2+\zt^2)^{-j/2} \cos(j\zf +\zd)
 \zJ \left (\zf\zH \tan^{-1} \frac{\zw}{\zk}\right ) .
 \label{keyint}\end{equation}
 We find that the inverse cosine transform of $P$ is
 \begin{equation}
\zr_{\zk,\text{per}} = 
 \frac{2L}{\pi} \sum_{n=1}^\zI \sum_{j=1}^n
(-1)^n \binom nj (-2\zk)^j 
 (\zw^2+\zt^2)^{-j/2}  \cos(2nL\zw+j\zf) 
\label{rhoper}\end{equation}
 and that of $B$ is
 \begin{equation}
\zr_{\zk,\text{bdry}} = 
 \frac2{\pi} \sum_{n=1}^\zI \sum_{j=2}^{n+1}
  c(n,j-1) \zk^{j-1}
   (\zw^2+\zt^2)^{-j/2}\cos(2nL\zw+j\zf). 
 \label{rhobdry}\end{equation}
 These formulas display $\zr_{\zk,\text{per}}$ and 
$\zr_{\zk,\text{bdry}}$  explicitly as sums over the periods $2nL$ of 
the periodic orbits.
 Finally, the complete representation of the density of eigenvalues is
 \begin{equation}
\rho = \zr_\text{N} +\zr_{\zk,\text{av}} + \zr_{\zk,\text{per}}   
+\zr_{\zk,\text{bdry}}   \,.
 \label{rhototal}\end{equation}

\emph{Remarks:} (1)
 Using (\ref{phi}) and trigonometric identities, all the 
terms in principle can be written as $\cos(2nL\zw)$ or $\sin(2nL\zw)$
 times rational functions of $\zw$ and $\zk$.
 (2) The series (\ref{rhoper}) and (\ref{rhobdry}) 
 are classically divergent.
   To what extent
  one should worry about this will be discussed in due course.

 \subsection{\label{ssec:locden}The local spectral density}

An alternative approach that has some advantages is to take the 
inverse cosine transform of (\ref{W}) (with $y=x$)
 \emph{before} integrating over~$x$.
 The intermediate result is a local spectral density.
The density resulting from (\ref{W1}) and (\ref{W2}) 
 is constant in~$x$
 (namely, $\frac1L(\zr_\text{N}+\zr_{\zk,\text{per}})$),  so 
nothing new happens there.
  The delta-function parts of  (\ref{W3}) and (\ref{W4})
  yield a local spectral density
 \begin{multline}
 \frac1\pi \left [\sum_{n=0}^\zI (-1)^{n+1}
 \cos\bigl( 2\zw(x-(n+1)L)\bigr) 
 +\sum_{n=1}^\zI (-1)^{n+1}\cos\bigl(2\zw(x+(n-1)L)\bigr) \right ]
 \\
= \frac1{\pi} \sum_{n=-\zI}^\zI (-1)^n \cos\bigl( 2\zw(x+nL)\bigr)
\label{nlocal}\end{multline}
 that is odd under $x\leftrightarrow L-x$, so that,
 as anticipated, the trace of this contribution vanishes by virtue 
of cancellation of the left-hand Neumann and right-hand Dirichlet 
effects (cf.\ Sec.~\ref{ssec:neum}). 
 The rest of the terms in (\ref{W3}) and~(\ref{W4}) 
 give the density 
 \begin{multline}
\frac1\pi \sum_{n=1}^\zI \sum_{j=1}^n  (-1)^{n-1} 
\binom nj (-2\zk)^j (\zw^2+\zk^2)^{-j/2} 
 \\
{}\zM \bigl[\cos \bigl( [2(n+1)L -2x]\zw +j\phi\bigr)
 + \cos \bigl( [2(n-1)L +2x]\zw +j\phi\bigr) \bigr],
 \label{bou-1}\end{multline}
 whose trace over $x$ is
 \begin{multline*}
  \zr_{\zk,\text{bou}} =
 \frac1{2\pi} \sum_{n=1}^\zI \sum_{j=1}^n (-1)^{n-1}
\binom nj \frac{(-2\zk)^j }{\zw}
 (\zw^2+\zk^2)^{-j/2} 
 \\
{}\zM \bigl[\sin\bigl(2(n+1)L\zw + j\zf\bigr) 
 -\sin\bigl(2nL\zw + j\zf\bigr)\bigr] 
 \end{multline*}
 \begin{multline}
 \hphantom{\zr_{\zk,\text{bou}}}
 +  \frac1{2\pi} \sum_{n=1}^\zI \sum_{j=1}^n (-1)^{n-1}
\binom nj \frac{(-2\zk)^j}{\zw}
 (\zw^2+\zk^2)^{-j/2} 
 \\
{}\zM \bigl[\sin\bigl(2nL\zw + j\zf\bigr) 
 -\sin\bigl(2(n-1)L\zw + j\zf\bigr)\bigr] ,
\label{boua} \end{multline}
 which ``obviously'' (but see Appendix~\ref{app:delta}
  and Secs. \ref{ssec:alt}--\ref{ssec:num}) simplifies to 
 \begin{multline}
 \zr_{\zk,\text{bou}} =
 \frac1{2\pi} \sum_{n=1}^\zI \sum_{j=1}^n (-1)^{n-1}
\binom nj  \frac{(-2\zk)^j}{\zw}
 (\zw^2+\zk^2)^{-j/2} 
 \\
 {}\zM \bigl[\sin\bigl(2(n+1)L\zw + j\zf\bigr) 
 -\sin\bigl(2(n-1)L\zw + j\zf\bigr)\bigr]. 
 \label{boub}\end{multline} 
 Presumably
 $\zr_{\zk,\text{bou}} =\zr_{\zk,\text{av}}+ \zr_{\zk,\text{bdry}}\,$,
though that is not obvious from the formulas (see Sec.~\ref{ssec:num}).

 \subsection{\label{ssec:alt}Alternative formulas}

 Still another periodic-orbit representation of the eigenvalue 
density is found in Appendix~\ref{app:Poisson}
  by working backwards from the 
transcendental equation determining the eigenvalues.
 One has  
 \begin{equation}
 \zr +\tfrac12 \zd(\zw) 
 \zH\zr_\text{Pois} = \zr_\text{Pois,per}+ \zr_\text{Pois,bou}
 \label{pois}\end{equation}
where 
 \begin{equation}
\zr_\text{Pois,per}(\zw) =
 \frac L\pi \left [ 1 + 2\sum_{n=1}^\zI \cos \bigl(2n(L\zw+\zf)\bigr) 
\right ],
 \label{poisper}\end{equation}
 \begin{equation}
\zr_\text{Pois,bou}(\zw) =
 \frac 1\pi\, \frac{\zk}{ \zw^2+\zk^2}
  \left [ 1 + 2\sum_{n=1}^\zI \cos \bigl(2n(L\zw+\zf)\bigr) \right ].
 \label{poisbou}\end{equation}
 Recall from (\ref{phi}) that
 \begin{equation}
\frac{\zk}{ \zw^2+\zk^2} = \frac{\sin\zf\cos\zf}\zw
  = \frac{\sin(2\zf)}{2\zw}\,,
 \label{B6}\end{equation}
so the prefactor in (\ref{poisbou}) could be combined with the phase 
shifts.
 Also, the delta function could be artificially written as a sum 
over the periodic orbits similar to~(\ref{A3}).
 Forgoing those possibilities, however, one observes that the sums in   
(\ref{poisper}) and (\ref{poisbou}) are identical except for a factor that 
depends only weakly on~$\zw$.
 Thus 
 \begin{equation}
 \begin{aligned}
 \zr_\text{Pois,per}(\zw)&= a(\zw) \zr_\text{Pois}(\zw),
\\ \noalign{\smallskip}
\zr_\text{Pois,bou}(\zw)&=\bigl(1-a(\zw)\bigr) 
\zr_\text{Pois}(\zw),
 \end{aligned} \zJ
\frac{1-a({\zw})}{a({\zw})} = \frac{\zk/L}{\zw^2+\zk^2}\,. 
\label{poisdecomp}\end{equation}

 Because of its $L$ dependence, one expects $\zr_\text{Pois,per}$
 to be the part of $\zr$ coming literally from the periodic orbits
 (as opposed to the closed orbits that reverse themselves and 
bounce off a nearby boundary), hence that
 \begin{equation}
\zr_\text{Pois,per} =\zr_\text{N} +\zr_{\zk,\text{per}}\,,
\label{poisrela}\end{equation}
\begin{equation}
\zr_\text{Pois,bou} +\tfrac12 \,\zd(\zw) =\zr_{\zk,\text{bou}} =
  \zr_{\zk,\text{av}}+\zr_{\zk,\text{bdry}}\,. 
 \label{poisrelb}\end{equation}
On the other hand, we should have
 \begin{equation}
\zr_{\zk,\text{bdry}} =
\zr_\text{Pois,bdry} \zH
 \frac 2\pi\, \frac{\zk}{ \zw^2+\zk^2}
  \sum_{n=1}^\zI \cos \bigl(2n(L\zw+\zf)\bigr). 
 \label{poisbdry}\end{equation}

The total, correct eigenvalue density $\zr$ does \emph{not} 
contain a delta function at $\zw=0$.
 Therefore, when $\zr$ is computed from the formulas of 
Sec.~\ref{ssec:eigden} 
 or from those of this section, a term 
$+\frac12\zd(w)$ must emerge from 
 the other terms.
 We previously observed that when $\zk=0$
the compensating delta comes from the other term of 
$\zr_\text{av}$ (\ref{rhoav}), properly interpreted.
 (Of course, in that case one can use Sec.~\ref{ssec:neum}
  and never introduce 
the more complicated expressions in the first place.)
 When (and only when) $\zk\ne 0$, the compensating delta must be 
hidden in the trigonometric sums, 
$\zr_{\zk,\text{per}} + \zr_{\zk,\text{bdry}}$ in the first approach
 and $\zr_\text{Pois,per} +\zr_\text{Pois,bdry}$ in the other.
 As Appendix~\ref{app:delta} shows, 
 the presence of such a term is a delicate 
question, since it can depend on the order of the terms in the 
 series.
 In contrast, the formulas of Sec.~\ref{ssec:locden}
  do not contain any delta 
functions, so one must not expect
 $\zr_{\zk,\text{per}} + \zr_{\zk,\text{bou}}$ to contribute any deltas 
in compensation.

 However, $\zr_{\zk,\text{bou}}$ remains somewhat ambiguous, because 
of the term-ordering issue. 
 The first term in (\ref{boua}) is the contribution from those 
pulses (\ref{W3}) that struck the right boundary, and the second 
term is from the pulses (\ref{W4}) that struck the left boundary first. 
 It is natural to define such sums by accumulating the terms in 
order of increasing path length (the same as the coefficient of 
$\zw$, or frequency of the spectral oscillations).
 In (\ref{bou-1}), however, that ordering depends on~$x$.
 There are three fairly reasonable things one could do:

 \begin{enumerate}
 \item Simplify (\ref{boub}) as it stands:
 \begin{equation}
\zr^\text{naive}_{\zk,\text{bou}} =
\frac1{\pi} \sum_{n=1}^\zI  \sum_{j=1}^n (-1)^{n-1}
\binom nj  \frac{(-2\zk)^j}{\zw} (\zw^2+\zk^2)^{-j/2}
 \sin(2L\zw )\cos(2nL\zw +j\zf).
\label{naive}\end{equation}
 Recall that (\ref{boub}) 
  pairs paths with the same number of bounces from 
the Robin boundary:  pulse~4 with pulse~7, 8~with 11, etc.

 \item Shift the index in the second term of (\ref{boua}) by one unit,
 and simplify:
\begin{equation}
 \zr^\text{pretrace}_{\zk,\text{bou}} =
 \frac1{\pi} \sum_{n=0}^\zI  \sum_{j=1}^{n+1} (-1)^n
\binom n{ j-1}  \frac{(-2\zk)^j}{\zw}  (\zw^2+\zk^2)^{-j/2}
 \sin(L\zw) \cos\bigl((2n+1)L\zw + j\phi\bigr).
\label{pretrace}\end{equation}
 Hereby paths with the same average length (as $x$ varies) are 
paired:  4~with~3, 8~with 7, etc.
 This choice seems to us to have the greatest physical 
justification.

 \item      In (\ref{boub})  shift an index by two
 units relative to the other, and simplify:
 \begin{equation}
\zr^\text{posttrace}_{\zk,\text{bou}} =  
 \frac1{2\pi} \sum_{n=0}^\zI  \sum_{j=1}^{n+1} (-1)^{n-1}
 \binom n{j-1}  \frac{2n-j+1}{n}
  \frac{(-2\zk)^j}{\zw}(\zw^2+\zk^2)^{-j/2} \sin(2nL\zw +j\zf)
\label{posttrace}\end{equation}
(with $(2n-j+1)/n = 0$ when $n=j-1=0$).
 Here all terms with the same frequency have been forcibly 
combined, so that the formula looks like a sum over periodic orbits 
alone.
(The labels ``pretrace'' and ``posttrace'' refer to the timing of 
the balancing of path or orbit lengths.  In both cases, unlike 
(\ref{rhobdry}), the cosine transform is performed before the 
trace.)

 \end{enumerate}

\begin{theorem}\label{th:specsumm}
The spectral implications of Theorem \ref{th:W} are summarized as 
follows, modulo terms supported at $\zw=0$ in the limit:
\begin{description}
 \item[\rm(a)] The local spectral density consists of the 
constant terms, (\ref{rhon}) plus (\ref{rhoper}) divided by~$L$,
arising from the periodic-orbit pulses (\ref{W1})--(\ref{W2}),
plus the terms (\ref{nlocal}) and (\ref{bou-1}) arising from the 
bounce-orbit pulses (\ref{W3})--(\ref{W4}).
\item[\rm(b)] The contribution of the periodic orbits to the 
density of eigenvalues is (\ref{rhon}) plus (\ref{rhoper}),
or alternatively  (\ref{poisper}) (the Poisson formula).
\item[\rm(c)] The contribution of the bounce orbits to the 
density of eigenvalues is (\ref{rhoav}) plus (\ref{rhobdry})
(trace-before-transform),
or alternatively any of (\ref{poisbou}) (Poisson),
(\ref{naive}), (\ref{pretrace}), or (\ref{posttrace}) 
(variants of transform-before-trace).
\end{description}
\end{theorem}

\subsection{\label{ssec:num}Symbolic and numerical evaluations}

 We resort now to \emph{Mathematica} \cite{mma}, which ``verifies''
 (\ref{poisrela}) and (\ref{poisbdry}) and furthermore reveals that
 (contrast (\ref{poisrelb}))
 \begin{equation}
\zr_\text{Pois,bou} =\zr^\text{posttrace}_{\zk,\text{bou}}\,.
 \label{surprise}\end{equation}
 That is, for any particular $n$ that we have tried, machine 
simplification shows the identity of the respective terms in each 
of these series pairs
 (but general~$n$ can't be handled).
 Also, numerical plots of the differences of respective partial sums 
show nothing but numerical noise.
 (The noise is usually very small 
 (machine-precision level)
 but not always: The sums over $j$ in 
such formulas as (\ref{rhobdry}) are numerically unstable when
  $\zw$ is very small.  See Fig.~\ref{fig:smallomega}(b).)
 On the other hand, partial sums of the three series 
 (\ref{naive})--(\ref{posttrace})
   are seen to be all different.

\begin{figure}
\includegraphics{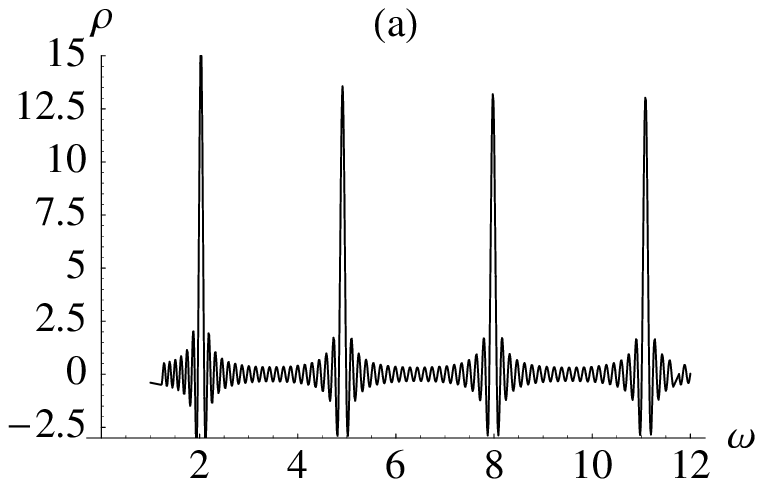}
\includegraphics{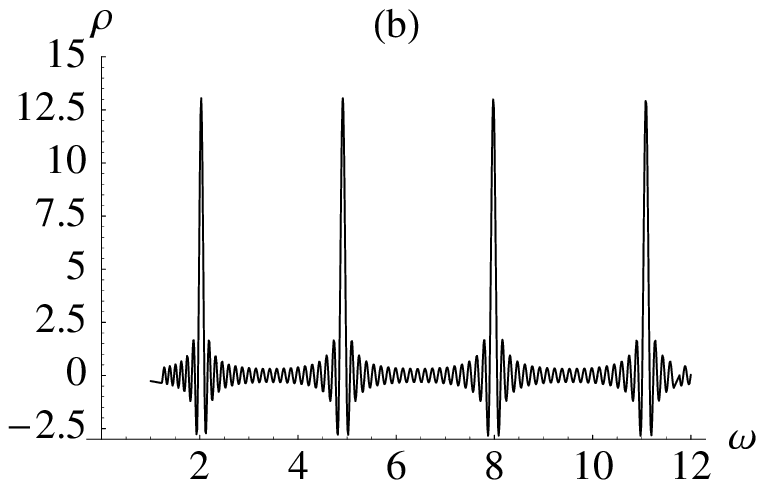}
\vskip 1truecm
\includegraphics{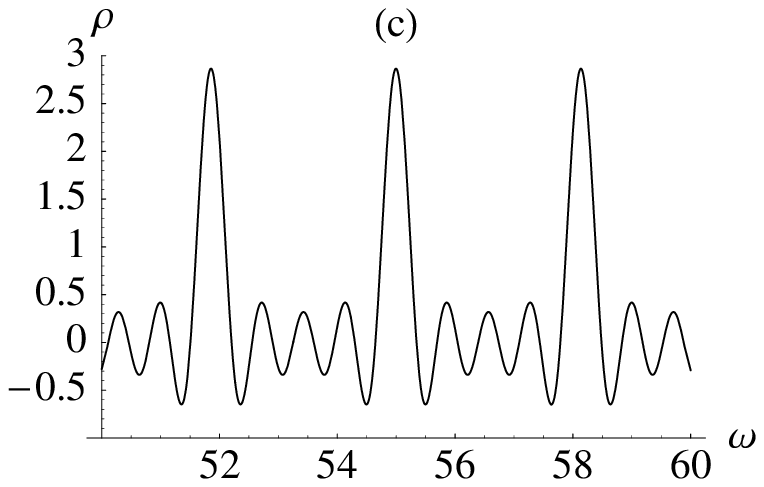}
\includegraphics{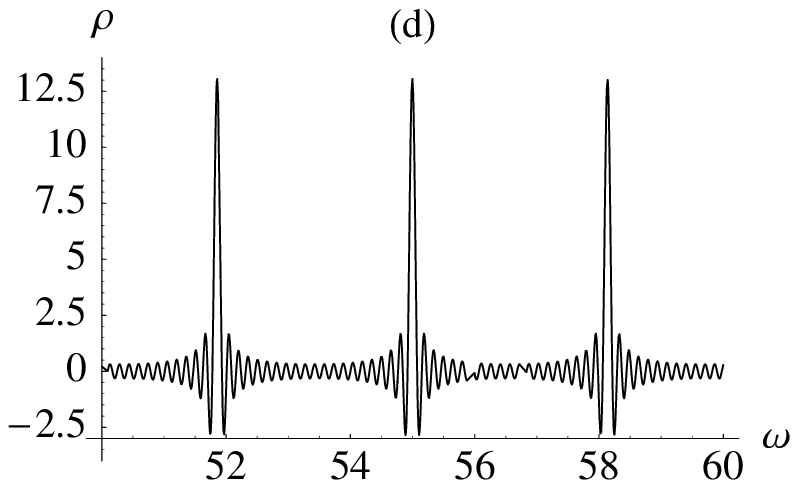}
\caption{(a) Eigenvalue density on interval $1<\zw<12$ computed 
from (\ref{pois}), partial sum $n_\text{max}=20$.
Other formulas [(\ref{rhototal}) and analogs built from 
(\ref{naive}), (\ref{pretrace}), and (\ref{posttrace})] 
give almost indistinguishable results.
(b) Same as (a) but from (\ref{poisper}), the contribution of 
genuinely periodic orbits only.  The plots are misleading 
because differences in peak height are more noticeable than 
differences in peak width.  Numerical integration confirms that 
the peaks in (a) have almost equal, unit strength, while those 
in (b) do not, in accordance with (\ref{B5}) and 
(\ref{poisdecomp}).
(c) Eigenvalue density on interval $50<\zw<60$ computed 
from (\ref{pois}) with $n_\text{max}=4$.
(d) Same as (c) but with $n_\text{max}=20$.
In all plots, $L=\zk=1$.} \label{fig:bigomega} 
\end{figure}

\begin{figure}
\includegraphics{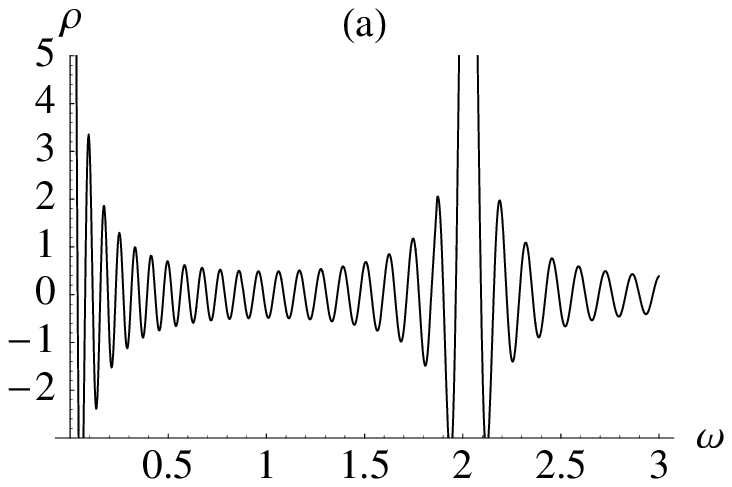}
\includegraphics{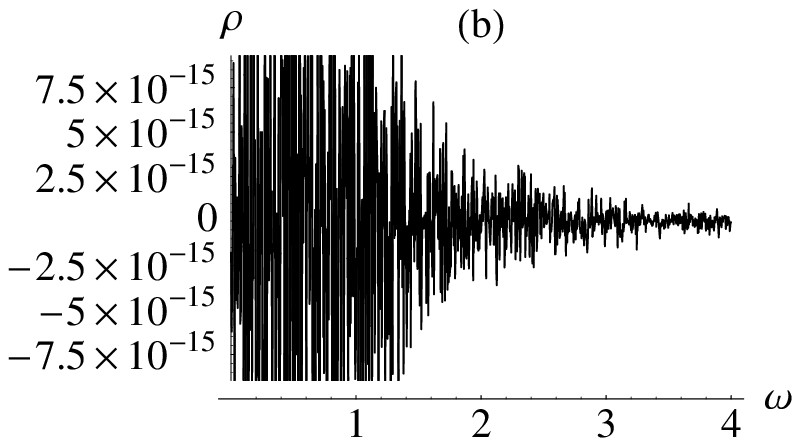}
\vskip 1truecm
\includegraphics{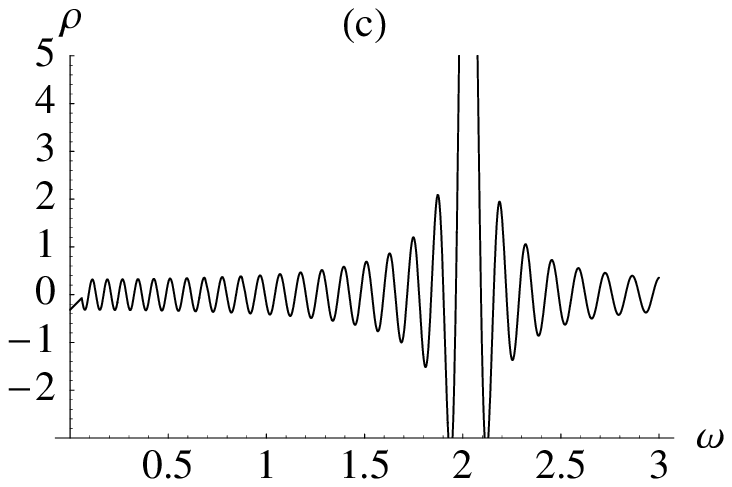}
\includegraphics{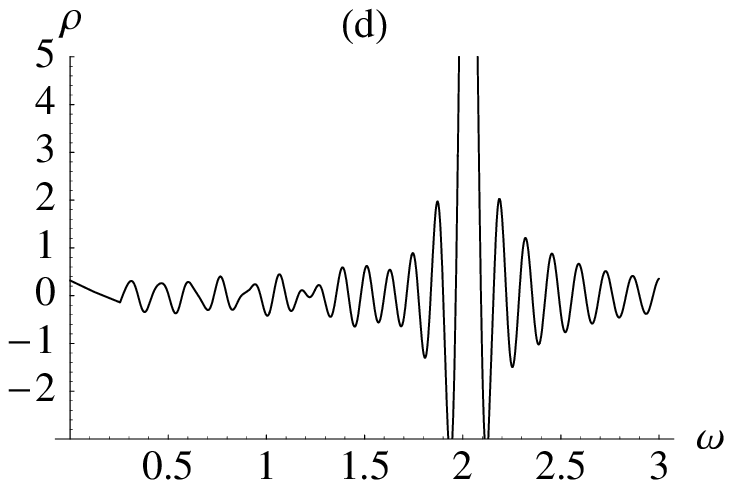}
\caption{(a) Eigenvalue density on interval $0<\zw<3$ computed 
from (\ref{pois}) (the Poisson formula) 
with $n_\text{max}=20$.  Cancellation of the 
delta function at $\zw=0$ is clearly visible.
(b) Difference between $\zr_{\zk,\text{bdry}}$ (\ref{rhobdry}) 
and $\zr_\text{Pois,bdry}$ (\ref{poisbdry}) on 
the interval $0<\zw<4$, $n_\text{max}=6$, 
showing only roundoff error.
(c)  Same as (a) but computed from 
the pathlength-balanced formula (\ref{pretrace})
[plus (\ref{rhon}) and (\ref{rhoper})].  The delta function is 
now missing.
(d) Same as (c) but computed from the ``naive'' 
formula (\ref{naive}); the difference 
presumably is  purely truncation error.
In all plots, $L=\zk=1$.}
\label{fig:smallomega}
\end{figure}

When plotted, all the series show, 
 even for very small partial sums,  the 
development of delta peaks at the (square roots of the) 
 eigenvalues of the Robin problem,
 \begin{equation}
-\, \od {^2y}{x^2}= \zw^2 y, \zJ
 y(L)=0, \zj \od yx(0) = \zk y(0).
 \label{eigp}\end{equation}
 In accordance with (\ref{poisdecomp}),
  the periodic terms alone already give the 
locations of the eigenvalues, and the bounce contributions are 
needed only to normalize the delta functions correctly.
 Examples are shown in Fig.~\ref{fig:bigomega}.
 For comparison, in the case $L=\zk=1$ we computed the 
lowest-lying
 eigenvalues by Newton's method to be
 \begin{equation}
\zw_1\approx 2.0288, \zj \zw_2\approx 4.9132 ,
\zj \zw_3\approx 7.9787, \zj \zw_4\approx 11.0855, 
 \label{loweig}\end{equation}
 and the large eigenvalues by perturbation theory to be
 \begin{equation}
\zw_n \approx (n-\tfrac12)\pi + \frac1{(n-\frac12)\pi}
 - \frac3{2(n-\frac12)^3\pi^3} + \frac3{(n-\frac12)^5 \pi^5}\,,
 \label{higheig}\end{equation}
with which (\ref{loweig}) already overlaps well.
 Fig.~\ref{fig:bigomega} 
 agrees well with these values.

We now turn to the issue of what happens for $\zw$ near $0$ 
 (Fig.~\ref{fig:smallomega}).
 The empirical evidence from the plots is that
  $\zr_{\zk,\text{per}}$ and $\zr_{\zk,\text{bdry}}\,$, 
 and equivalently $\zr_\text{Pois,per}$ and $\zr_\text{Pois,bdry}\,$,
 are behaving exactly as expected:
The right-hand half of a delta function is building up at the 
origin to cancel the (correct, but misleading) 
 $-\frac12\zd(\zw)$ in the formulas of
 Secs.~\ref{ssec:eigden} and~\ref{ssec:alt}.
Because of (\ref{surprise}), we get the same behavior from 
 $\zr_{\zk,\text{per}}$ and $\zr^\text{posttrace}_{\zk,\text{bou}}\,$;
 that is, the forcible reinterpretation of contributions to the 
local spectral density from closed but nonperiodic orbits 
as if they were contributions from periodic orbits has, by the 
mathematical phenomenon of Appendix~\ref{app:delta}, 
 introduced a truly spurious 
 term $+\frac12\zd(\zw)$ into the formulas of 
Sec.~\ref{ssec:locden}!
 On the other hand, both $\zr^\text{pretrace}_{\zk,\text{bou}}$
 and $\zr^\text{naive}_{\zk,\text{bou}}$ are converging to the 
right answer for the context of Sec.~\ref{ssec:locden}.
  Their partial sums differ,
 but the difference (apparently) goes away in the limit.
 These series are yielding negative delta functions that precisely 
cancel the delta behavior of $\zr_{\zk,\text{per}}\,$,
 so that no  $\zd(\zw)$ is ever visible in the total density
 (see Figs.~\ref{fig:smallomega}(c,d)).

\subsection{\label{ssec:impl}Broader implications}

 This simple one-dimensional model is trivial in comparison with 
many problems treated by similar methods in the contemporary 
literature.
 It cannot manifest quantum chaos;
 it doesn't require stationary-phase approximations;
 it isn't even relevant to the sort of semiclassical approximation 
that becomes necessary when a potential is added to the
  one-dimensional Schr\"odinger equation (\ref{eigp}) 
  \cite{lgacee3}.
Nevertheless, it is instructive.
 The mere fact that so much information can be obtained exactly 
 means that the model can be understood in complete detail and 
stands as a benchmark against which partial and approximate 
solutions of more complicated models can be appraised.

 In particular, the delicate problem of the delta function at 
$\zw=0$ is likely to have broader implications.
 In more complicated problems such precise analysis at the bottom 
end of the spectrum is usually lost in the noise of the 
semiclassical and stationary-phase approximations.
Periodic-orbit reconstructions of spectra are observed to be 
surprisingly good at low frequencies, except \emph{very} close to 
$0$, where a spurious peak often occurs.
 (Look at Fig.~17 of \cite{BB3} and Fig.~29 of~\cite{FW}.)
 Our analysis suggests that this phenomenon represents not a 
breakdown of the semiclassical approximation so much as an 
ambiguity in the ordering of the terms in the badly convergent 
periodic-orbit sum.

 A related point is the significance of contributions from orbits 
that are closed but not periodic.  Usually such contributions are 
discarded in the process of stationary-phase approximation.
 However, they are needed to get the boundary terms in 
$\zr_\text{av}$ \cite{BB1,SSCL}, and Jaffe and Scardicchio 
\cite{JS,SJ}
 have recently emphasized their importance in calculations of total 
vacuum energy.
 As semiclassical calculations are carried beyond the lowest order 
in~$\hbar$, these orbits will need to be included, and one must 
grapple with the question of whether they must be kept separate,
 as in (\ref{boua}), or 
will be effectively absorbed into the periodic contributions as in 
$\zr_{\zk,\text{bdry}}$ (\ref{rhobdry}) and 
 $\zr^\text{posttrace}_{\zk,\text{bou}}$ (\ref{posttrace}).
In our model they did not influence the location of the eigenvalues, but
they were needed for proper normalization;
 is there some broader significance to that?

 \begin{acknowledgments}
 We thank Ricardo Estrada, Klaus Kirsten, and Peter Kuchment for 
helpful remarks. JDB gratefully acknowledges the support of the 
Advanced Technology Program of the
State of Texas,  Award  010366-0218-2001 
to Robert Kenefick, and we thank Dr.\ Kenefick for encouraging us 
to work together.
  \end{acknowledgments}

\appendix 

 \section{\label{app:delta}Nonabsolute convergence of oscillatory
distributions}

 In the periodic-orbit approach to spectral theory,
  sums of the type
 \begin{equation}
\sum_{n=0}^\zI \sum_{m=1}^M a_m(n,\zw) \cos(n\zw +\zd_m)
 \label{A1}\end{equation}
 are often encountered.
 Typically $M$ is a fixed, small number (such as $2$), and 
$a_m(n,\zw)$ varies slowly with $\zw$ but may become singular as 
$\zw\to0$ and, most important, does not approach $0$ as $n\to\zI$. 
 Such a series does not converge at all according to the 
definitions of classical analysis, but nevertheless it may converge 
in the sense of distributions.
 The question then arises whether the distributional limit can depend 
on the order in which the terms are added,
 in analogy with numerical series that are only conditionally 
convergent.
 We consider here distributions defined on the nonnegative real 
line.

\begin{theorem}\label{th:distconv}
   In (\ref{A1}) assume that $M$ is independent 
of~$n$ and that the functions $a_m(n,\zw)$ are smooth ($C^\zI$) for
 $0<\zw<\zI$ and they and all their 
derivatives are uniformly polynomially bounded as $n\to\zI$.
 Then any two orderings of the terms in $(A1)$ define
distributions that 
 coincide on test functions with support in the interior 
 (i.e., $\zf(\zw)=0$ in a neighborhood of the origin).
 As distributions on test functions defined in $[0,\zI)$
 they may
 differ  by a distribution supported at $\zw=0$ (necessarily a linear 
combination of $\zd(\zw)$ and its derivatives).
\end{theorem}

\emph{Proof:}  The definition of distributional convergence 
  is that for any test function $\zf$, 
 \begin{equation}
\sum_{n=0}^\zI \sum_{m=1}^M \int_0^\zI a_m(n,\zw) \cos(n\zw +\zd_m)
 \zf(\zw)\,d\zw  
 \label{A2}\end{equation}
 converges in the classical sense
 and defines a continuous linear functional of~$\zf$.
 If $\zf(\zw)=0$ in a neighborhood of the origin, then repeated 
integration by parts,
 \begin{eqnarray*}
 \int_0^\zI a_m(n,\zw) \cos(n\zw +\zd_m) \zf(\zw)\,d\zw 
 &=&
\int_0^\zI \frac1{n^{4p}}\,\pd{^{4p}}{\zw^{4p}}\cos(n\zw +\zd_m)
a_m(n,\zw) \zf(\zw)\,d\zw
 \\
 &=&  \frac1{n^{4p}}\int_0^\zI \cos(n\zw +\zd_m)
\pd{^{4p}}{\zw^{4p}}[a_m(n,\zw) \zf(\zw)]\,d\zw,
 \end{eqnarray*}
  shows that each integral in (\ref{A2}) falls off 
faster than any power of~$n$.
Taking $p$ sufficiently large guarantees that (\ref{A2})
 converges absolutely   and hence can be reordered at 
will.
 If the support of $\zf$ includes the origin, 
 the endpoint terms from the partial integrations may converge to a 
well-defined delta-type distribution.  
 If not, it may still be possible to extend the distributions  to 
such test functions by ``regularization'' \cite{EF};
 in that case the coefficients in the delta sum are somewhat 
ambiguous.

\emph{Example:} Consider
 \begin{equation}
\sum_{n=0}^\zI \frac1{\zw}\, [\sin((n+1)\zw) - \sin(n\zw)]. 
 \label{A3}\end{equation}
 From one point of view, shifting the index in the first term 
yields
 \[
\sum_{n=0}^\zI \frac1{\zw}\, [\sin(n\zw) - \sin(n\zw)] =0.
 \]
 On the other hand, the $n$th partial sum of (\ref{A3}) as written is
 \[
 \sum_{n=0}^{N-1} \frac1{\zw}\,[\sin((n+1)\zw) - \sin(n\zw)] 
 = \frac1{\zw}\sin(N\zw),
 \]
 and the latter converges to $\frac{\pi}2 \zd(\zw)$:
 \begin{eqnarray*}
\int_0^\zI \frac1{\zw}\sin(N\zw) \zf(\zw)\,d\zw 
&=& \int_0^\zI \frac1{\zw}\sin(N\zw) [\zf(0) + O(\zw)]\,d\zw
 \\
& =& \zf(0) \int_0^\zI \frac1{z}\sin(z) \, dz + \int_0^\zI \sin(N\zw) 
O(1)\, d\zw;
\end{eqnarray*}
 the second term vanishes as $N\to\zI$ by the Riemann--Lebesgue 
theorem, and the integral in the first term equals $\frac{\pi}2$
\cite[(3.721.1)]{GR}.

\emph{Remark:}  That the series rearrangements
 in the body of the paper obey the 
polynomial boundedness requirement in Theorem~\ref{th:distconv}
  follows from the bound \cite[(2.14.13)]{AS}
 \begin{equation}
|L^1_{n-1}(x)| \, e^{-x/2} \le n \zj(x\ge0)
 \label{lagbound}\end{equation}
on the Laguerre polynomial.
 Individual terms in (\ref{boub}) (for instance) 
 with $j\approx \frac n2$ 
 can grow exponentially with~$n$, but we have never realigned terms 
with differing values of~$j$.

 \section{\label{app:Poisson}Reconstruction of the periodic-orbit 
sum  by Poisson summation}

 The following  is a close analogue of a two-dimensional 
calculation in Sec.~2.2 of \cite{SPSUS}.
 It requires knowledge of the eigenvalue condition (\ref{B1})
  (but not of explicit formulas for its solutions).

 The square roots of the 
 eigenvalues of the Robin--Dirichlet problem (\ref{eigp}) with 
 $\zk>0$ are the positive roots of
 \begin{equation}
\tan(L\zw ) = -\, \frac{\zw}{\zk} \,,
 \label{B1}\end{equation}
 which may be parametrized as
 \begin{equation}
k\pi = L\zw  +\zf, \zJ  \zf\zH \tan^{-1} \left(\frac{\zw}{\zk}\right),
 \zj k=1,2,\ldots\,.
 \label{B2}\end{equation}
One can view (\ref{B2}) as giving $k$ as a function of $\zw$,
  which  naturally extends to the whole real line as a monotonic 
  and odd function with derivative
 \begin{equation}
\od k{\zw} =\frac1{\pi}\left( L + \frac{\zk}{\zw^2+\zk^2} \right ).
 \label{B3}\end{equation}
 Thus $\zw_k$ is defined for all integers, with $\zw_{-k}= \zw_k$
 and $\zw_0=0$ (which is not an eigenvalue).
 So the eigenvalue density is
 \begin{equation}
\sum_{k=1}^\zI \zd(\zw-\zw_k)
  = \frac12 \sum_{k=-\zI}^\zI \zd(\zw-|\zw_k|) -\frac12\zd(\zw). 
 \label{B4}\end{equation}
 
 By the Poisson summation formula \cite{poisref}
 (and (\ref{B3}) and (\ref{B2})) one has
 \begin{eqnarray*}
\frac12 \sum_{k=-\zI}^\zI \zd(\zw-|\zw_k|) &=&
\frac12 \sum_{n=-\zI}^\zI \int_{-\zI}^\zI dk\,     \zd(\zw-|\zw_k|)
 e^{2\pi ikn} 
 \\
 &=& \frac12\sum_{n=-\zI}^\zI \int_{-\zI}^\zI d\tilde{\zw} \, \od 
k{\tilde{\zw}} \, \zd(\zw-|\tilde\zw|) e^{2\pi ikn}  
 \\
 &=&\zy(\zw)\sum_{n=-\zI}^\zI 
 \frac1{\pi}\left ( L + \frac{\zk}{\zw^2+\zk^2} \right )
 \cos\bigl(2n(L\zw +\zf)\bigr),
\end{eqnarray*}
since $\zw=|\tilde\zw|$ has two roots $\tilde\zw$ for $\zw>0$ and 
 none for $\zw <0$.
Therefore, 
 \begin{equation}
\sum_{k=1}^\zI \zd(\zw-\zw_k) =
-\, \frac12 \zd(\zw) +\frac{\zy(\zw)}{\pi}\left [ 
L + \frac{\zk}{\zw^2+\zk^2} 
 +2\sum_{n=1}^\zI \left (L + \frac{\zk}{\zw^2+\zk^2}  \right ) 
 \cos \bigl( 2n(L\zw +\zf)\bigr)\right ]. 
 \label{B5}\end{equation}

 It is natural to associate the terms in (\ref{B5})
  with prefactor $L$ 
 directly with the periodic orbits and to regard the other terms 
as the traces of the contributions of the 
bounce orbits.
 With this interpretation, and confining attention to
$\zw\ge0$,
 we have arrived at (\ref{pois})--(\ref{poisbou}).

\goodbreak

\end{document}